*Article*

# A Repeated Game Freeway Lane Changing Model

**Kyungwon Kang [1] and Hesham A. Rakha [1,*]**

[1] Center for Sustainable Mobility, Virginia Tech Transportation Institute, Blacksburg, Virginia, 24061, United States

* Correspondence: hrakha@vt.edu; Tel.: +1-540-231-1505

**Abstract:** Lane changes are complex safety- and throughput-critical driver actions. Most lane-changing models deal with lane-changing maneuvers solely from the merging driver's standpoint and thus ignore driver interaction. To overcome this shortcoming, we develop a game-theoretical decision-making model and validate the model using empirical merging maneuver data at a freeway on-ramp [1-2]. Specifically, this paper advances our repeated game model in [2] by using updated payoff functions. Validation results using the Next Generation SIMulation (NGSIM) empirical data show that the developed game-theoretical model provides better prediction accuracy compared to previous work, with correct predictions approximately 86% of the time. In addition, a sensitivity analysis demonstrates the rationality and sensitivity of the model to variations in various factors. To provide evidence of the benefits of the repeated game approach, which takes into account previous decision-making results, a case study is conducted using an agent-based simulation model. The proposed repeated game model produces superior performance to a one-shot game model, when simulating actual freeway merging behaviors. Finally, this lane change model, which captures the collective decision-making between human drivers, can be used to develop automated vehicle driving strategies.

## 1. Introduction

Driving behavior strongly affects the safety and throughput of the transportation system [3]. Due to its interference with surrounding vehicles, lane-changing significantly affects traffic stream flow. Several studies have concluded that lane-changing produce a capacity drop forming a bottleneck [4-6]. The impacts of lane-changing maneuvers have been modeled in several studies [7-10]. In particular, Liu et al. [11] argued that traffic conflicts between merging and through vehicles, which are common near freeway on-ramps, are notable for inducing shockwaves, resulting in congestion. In order to analyze traffic flow, therefore, development of a state-of-the-art lane-changing model is important.

The applications of lane-changing models can be broadly classified into two groups: adaptive cruise control and microscopic traffic simulation [3]. Driving assistance models for adaptive cruise control consist of collision prevention models and automation models [12]. In addition, driving decision models focus on drivers' lane changing decisions for different traffic conditions and for different situational and environmental characteristics [12]. Lane-changing models were proposed based on various methodologies, which are reviewed in the next section, and calibrated based on field data collected on freeways. These models are an important component of microscopic traffic simulation [13]. Most models, however, focus on only the lane-changing vehicle in the decision-making and vehicle control, which could be detrimental in microscopic traffic simulation, as interaction with surrounding vehicles is also critical in lane-changing. Specifically, drivers of vehicles surrounding the lane-changing vehicle, especially the closest following vehicle in the target lane, react after recognizing a lane-changing vehicle's intention to change lanes. For example, a human driver will sometimes not allow a lane change. Even though this type of competitive lane-changing behavior is rarely observed, decision-making considering drivers' interaction when changing lanes should be studied in order to develop a precise lane-changing model.

In addition, modeling a driving strategy for automated vehicles (AVs) gives rise to a new application for lane-changing models. The introduction of AVs onto the roadway means that reasonable lane-changing decision-making can be conducted by an intelligent robot or well-programmed machine. During the transition to fully autonomous transportation systems, harmonization with human drivers will be necessary for the operation of AVs. Therefore, development of a realistic lane-changing model that can depict human drivers' decision-making is also required to enhance AVs' driving performance.

This paper enhances our repeated game lane-changing model proposed in [2] and evaluates the proposed model's performance. The paper begins by introducing the lane-changing models based on various methodologies, including a game theoretical approach. To enhance model efficiency and complement the multivariate function in the previous model, the payoff functions for a stage game are reformulated in Section 3. This study also applies the repeated game approach, which uses cumulative payoffs, in order to capture realistic human driver behavior at a freeway merging section. Both the repeated game model and the one-shot game model based on the reformed stage game are calibrated and validated using empirical data extracted from the NGSIM dataset [14-15] to demonstrate the prediction ability. In the rest of this paper, we present a sensitivity analysis to describe the stage game's efficiency. The simulation case study using an agent-based model (ABM) follows. Finally, we draw concluding remarks on this work, and point out areas of potential future research.

## 2. Literature Review

A comprehensive literature review is required to introduce previous research efforts and present the motivations for this study. This section begins with a review of lane-changing models, focusing on methodologies. Then, game theory-based models are introduced in detail. Based upon the literature review, the motivations for the study are presented.

### 2.1. Lane-Changing Decision-Making Models

In general, the lane-changing process can be categorized as a sequence of four steps: (1) checking for lane change necessity, (2) lane selection to decide on a target lane, (3) gap choice in the target lane, and (4) lane-changing execution through gap acceptance. To model lane-changing behaviors, lane-changing models have been developed using various methodologies that can be grouped into four types: (1) rule-based models, (2) discrete-choice-based models, (3) artificial intelligence models, and (4) incentive-based models [3].

The first model type, the rule-based model, is one of the most popular driver-perspective based methodologies [3]. Drivers' decisions in the lane-changing process are simply defined as the independent variable. Gipps [16] initially introduced a lane-changing model covering various urban driving situations, which was intended for microscopic traffic simulation tools [17]. Gipps' model represented the lane-changing process as a decision tree with a series of fixed conditions, where the final output of this rule-based triggered event is a binary choice (i.e., change or not change) [3]. The CORSIM model classified lane changes into two types: (1) discretionary lane-changing (DLC), which occurs when a driver is unsatisfied with the driving situation in their current lane, while the target lane shows better driving conditions; (2) mandatory lane-changing (MLC), which is coercively required according to the route choice (i.e., lane change toward on-ramp or off-ramp) [18-19]. Rahman et al. [3] categorized the game theory-based model, which explains lane-changing when a traffic conflict arises between the merging vehicle and the closest following vehicle in the target lane, as a rule-based model. Game theory, which is used in this paper, is the study of mathematical models of conflict and cooperation between decision-makers [20]. It focuses on decision-making in consideration of the interaction between intelligent drivers. Using a game theoretical approach is advantageous in that it takes into account the behaviors of a following vehicle driver in the target lane, while the other approaches introduced above focus only on the lane-changing vehicle driver's decision.

The second model type, the discrete-choice model, relies on a logit or probit model to describe lane-changing maneuvers. Lane-changing is decided based on probabilistic results instead of binary answers. Ahmed [21] modeled lane-changing motivation (i.e., trigger to change a lane), target lane choice, and gap acceptance, presenting three categories of lane-changing: DLC, MLC, and forced merging (FM), in which a gap is not sufficient but a driver nonetheless executes a lane-changing maneuver in heavily congested traffic conditions. Ahmed [21] assumed that critical gaps follow a lognormal distribution to guarantee that they are nonnegative. Toledo et al. [22] developed a probabilistic lane-changing decision model by combining MLC and DLC through a single utility function. Both models developed by Ahmed [21] and Toledo et al. [22] considered drivers' heterogeneity, such as aggressiveness and driving skill level, using a random term as one of the explanatory variables.

The third model type, which includes fuzzy models and artificial neural network (ANN) models, are artificial intelligence models. The fuzzy model considers humans' imprecise perception and decision biases, and incorporates more variables than the common mathematical models [23]. However, the fuzzy model has disadvantages, such as unexpected difficulties and complexity in the fuzzy rules [23]. The ANN model processes information using functional architecture and mathematical models that are similar to the neuron structure of the human brain [3]. Hunt and Lyons [24] modeled the lane-changing decisions of drivers on dual carriageways. Since the neural network model is completely data-driven and requires field-collected traffic data, Hunt and Lyons used interactive driving simulation to train the model. As this example shows, one major disadvantage of the ANN model is that it requires a huge amount of data to be optimized and also requires a training period.

The last type of model, the incentive-based model, models lane-changing desire utilizing the defined incentive. In other words, this model assumes that a driver chooses to change lanes in order to maximize their benefits [3]. The minimizing overall braking induced by lane change (MOBIL) model, which was developed in Kesting et al. [13], is based on measuring both the attractiveness and the risk associated with lane changes in terms of accelerations. Therefore, both the incentive criterion and the safety constraint are formed using the acceleration function of the underlying car-following model. In addition, the model attempts to capture the degree of passive cooperation among drivers, using the politeness factor as a weight on the term for total advantage of the surrounding vehicles.

## 2.2. Game Theory-based Lane-Changing Decision-Making Model

It is clear that lane-changing involves not only a driver of the subject vehicle (SV), who is motivated to change lanes, but also a driver of the lag vehicle (LV) in the target lane, who controls their own vehicle (i.e. the LV) after perceiving the lane-changing vehicle in the adjacent lane. Specifically, the driver of SV controls their longitudinal and lateral movements to safely change a lane in consideration of surrounding vehicles, and the driver of the LV responds by showing acceptance or non-acceptance of an SV's lane-changing intention. This decision-making process involving both drivers motivated previous studies to use a game theoretical approach. Game theory-based models, therefore, were modeled as a two-player non-cooperative game.

Kita [25] modeled merging-giveway interaction between vehicles in a merging section based on a game theoretical approach. The action strategies of the driver of SV are merging or maintaining the current lane, while the strategies of the driver of LV in the target lane are giving way (i.e., yielding) or not. Kita [25] modeled interaction between drivers as a game under perfect information conditions. However, perfect information in game theory indicates that all players have perfect and instantaneous knowledge of their own utility and the events that have previously occurred. In a traditional transportation environment, in which a driver becomes aware of surroundings through sight only, this assumption is irrational. Additionally, Kita's model assumed that vehicle speeds were constant during the merging process, which is likewise unrealistic [11].

Liu et al. [11] modeled merging and yielding behavior using modeled payoff functions about the drivers' objectives. In Liu et al. [11], the objective of the driver of SV is to minimize the time spent in an acceleration lane subject to safety constraints, while the objective of the driver of LV is to minimize speed variation. The payoffs of drivers of the SV and LV were formulated using acceleration level and time that the merging vehicle spends in the acceleration lane for each action strategy, respectively. However, the driver of SV occasionally showed different behaviors, which were assumed to be based on the objective of the driver of SV. Kondyli and Elefteriadou [26] found that all drivers want to reach a speed close to the freeway speed or the speed limit, if there is no lead vehicle. This speed synchronization process that causes drivers to accelerate when arriving at the beginning of an acceleration lane was observed at a merging section on a freeway [27]. To solve the game, Liu et al. [11] proposed a bi-level calibration framework, in which the upper level programming is an ordinary least square problem and the lower level programming is a linear complementarity problem for finding the Nash equilibrium.

In [1], we modeled a decision-making game model for merging maneuvers using five decision factors and evaluated the proposed model using NGSIM data. In addition, we introduced a repeated game approach in order to avoid instantaneous fluctuation of decisions in microscopic simulation [2]. Even though these models showed high prediction accuracy, there were limitations, namely: the number of data showing all action strategies sets was unbalanced due to data collection during the morning peak time, and the model validation results were unable to show the distinct performance of the repeated game approach in microscopic simulation.

The development of advanced vehicle technologies (e.g., vehicle-to-vehicle communication) and AVs, has led recent research efforts to focus on the cooperative interaction between vehicles [28-29]. Talebpour et al. [29], for instance, modeled both mandatory and discretionary lane-changing by applying the Harsanyi transformation [30] within a connected environment. And Yu et al. [31] designed a human-like, game theory-based controller for AVs in consideration of mixed traffic.

*2.3. Motivation and Contribution of the Paper*

The following are the contributions of the paper. First, we enhance the payoff functions that were previously developed in [1-2] by taking into consideration multiple decision factors and normalizing the decision variables. Multivariate functions using variables, which have different units, may induce a trivial equilibrium solution when variables are correlated. To solve this issue, we reformulated the payoff function by considering dimensionless variables. Second, we validate and compare the previous and proposed models. Third, we conduct a sensitivity analysis of the proposed model performance. Fourth, we demonstrate the benefits of a repeated-game approach using a simulation tool. The repeated game model first introduced in [2], in which a stage game is repeatedly played taking into consideration previous game results, showed no evidence of benefits compared to a one-shot game model played independently based on instantaneous data at every decision point. If there is competition between drivers due to an ambiguous merging situation—for example, not only small lag spacings but also similar vehicle speeds—, the one-shot game model may be sensitive to instantaneous data, causing fluctuations in driver decisions during the decision-making process. On the other hand, the repeated game model's initial cooperative decision can be expected to remain the same when there is only a slight variation in payoffs. Furthermore, the game model can produce a change from a non-cooperative to a cooperative game. Even though this type of driver competition in merging seldom occurs, the robust game model can be integrated into a microscopic traffic simulation software in order to simulate stereotypical vehicle movement patterns. Consequently, in this study we adopt the previous repeated game approach with enhancements in the payoff function and then provide evidence of the repeated game model's benefits through a case study.

Lastly, a desired acceleration level, which is calculated to achieve the action set chosen by both players, should be an additional component of a vehicle acceleration model. A lane-changing model based on a game theoretical approach captures the decision-making process between two intelligent decision-makers. The model output is an action that will be conducted by two players at future time steps rather than a decision to start lane-changing. To depict practical lane-changing behaviors in a microscopic traffic simulator, therefore, the game model should be integrated with other models, such as car-following, lane selection, and gap acceptance models. This study develops an A simulation model based on ABM, including a vehicle acceleration controller based on the game model and a car-following model, then conducts a simulation study to evaluate the performance of the repeated game model.

## 3. Merging decision-making model using a repeated game concept

As previously noted, this study aims at developing a decision-making game for merging maneuvers on a freeway based on the repeated game concept. The following subsections describe, in detail, a stage game for merging decision-making and repeated game design and the development of the player payoff functions.

### 3.1. Stage Game Design

The game model defines the number of players, action strategies of each player, and corresponding payoff functions to describe the outcome for each player throughout the game [32]. This study adopts the decision-making game model structure for merging maneuvers proposed by the authors in 2017, which consists of two players: the drivers of the SV and LV. The driver of SV, who wants to make a lane change, has three action strategies (see Figure 1(a)): (1) change lane ($s_1$), (2) wait for the LV's overtaking maneuver ($s_2$), or (3) overtake the LV and use a forward gap to merge ($s_3$). The opposite player, the driver of LV, has two action strategies (see Figure 1(b)): (1) yield to allow the lane change maneuver of the driver of SV ($l_1$) or (2) block the SV's merging maneuver by decreasing the spacing available for the SV ($l_2$) [1]. In real life situations, the driver of LV can choose lane-changing to the left lane to avoid potential collision or considerable deceleration [33], and this lane-changing behavior was considered as an action strategy of the driver of LV in [29]. Freeway vehicles on the rightmost lane generally change lanes from the rightmost lane upstream of the merging section after perceiving the approach of the merging vehicle in order to maintain their speed. Since this mainline vehicle's lane change is conducted earlier and thus does not involve interaction with the merging vehicle, this study does not include a lane-changing action as one of the actions of the driver of the LV in the proposed merging game.

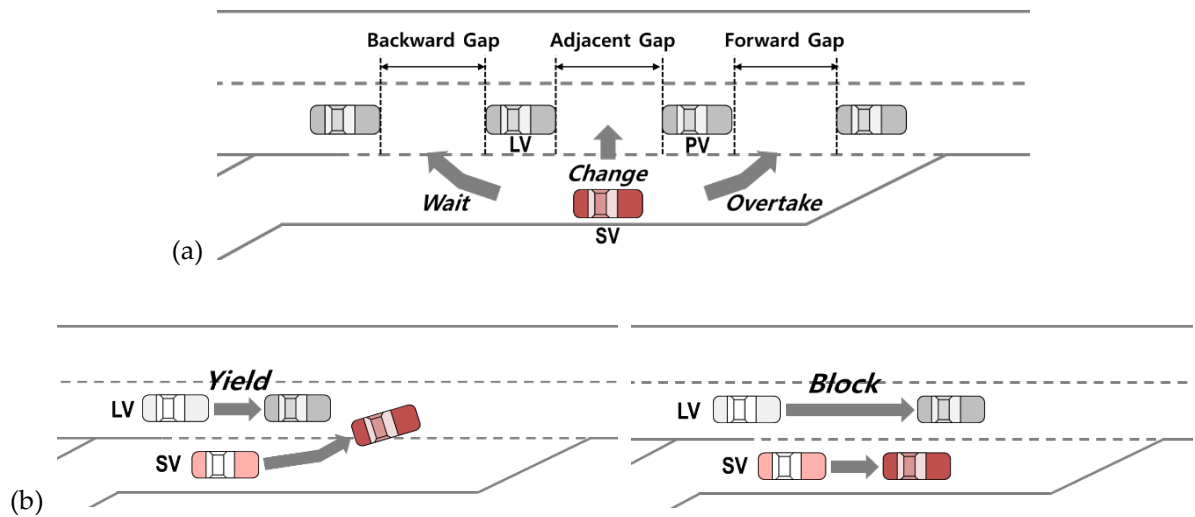

**Figure 1.** Players' strategies for merging maneuver: (a) the driver of SV; (b) the driver of LV ([1]).

Let $S = \{s_1, s_2, s_3\}$ and $L = \{l_1, l_2\}$ denote the set of pure strategies for the drivers of the SV and LV, respectively. In addition, $a = (s_i, l_j)$ denotes a set of actions ($a \in S \times L$) where $i$ and $j$ indicate the index of action strategies of the drivers of the SV and LV (i.e., $i$ = 1, 2, 3 and $j$ = 1, 2). As such, a total of six sets of action strategies were defined for the non-cooperative decision-making stage game. In these action strategies, $(s_1, l_1)$, $(s_2, l_2)$, and $(s_3, l_1)$ are cooperative action strategies, whereas both $(s_1, l_2)$ and $(s_2, l_1)$ are non-cooperative strategies in which both players' compete to achieve their objectives. The action strategy $(s_3, l_2)$ is neither cooperative nor competitive. The proposed stage game with imperfect information, which captures the fact that players are simply unaware of the actions chosen by other players, is represented in Figure 2. In the figure, a dashed line uniting three nodes, which implies imperfect information, indicates that the players do not know which node they are in. This means that there is no sequence in making a decision, and thus the driver of LV does not know the SV's movement. Moreover, $P_{ij}$ and $Q_{ij}$ denote the payoff for the drivers of the SV and LV for each action strategy $a_{ij}$, respectively.

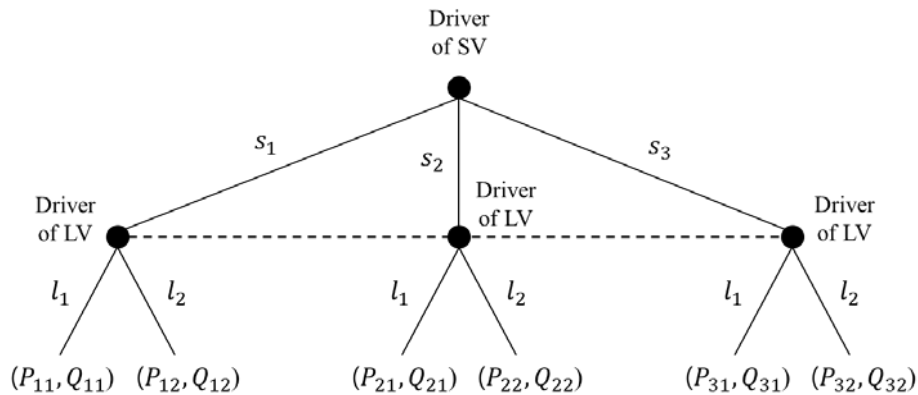

**Figure 2.** Merging decision-making game in the extensive form.

The drivers initially play the stage game to decide on an individual action at the moment when a SV, a LV, and a preceding (lead) vehicle (PV) are identified ([1]). It was assumed that the initial game is played when the driver of the SV reaches the start of an acceleration lane. Additional stage games are formed by overtaking the PV or waiting to be overtaken by the LV. In other words, the stage game is re-built when a change of surrounding vehicles occurs, i.e., PV or LV, in the target lane.

### 3.2. Repeated Game Design

In the game model, one of the characteristics to be specified is the number of games to be repeated [25]. Aside from [2], to the author's knowledge none of the other game theory-based models used the repeated game approach. In the authors' previous study, a repeated game approach was used in order to depict a practical decision-making process for merging maneuvers. In real life, at a freeway merging section in a traditional transportation environment, a driver continuously makes a decision using the data taken in by sight and controls the vehicle to fulfill their decision. When the merging vehicle enters the acceleration lane, the driver of the SV selects a gap type to change a lane and then directs their vehicle accordingly. The driver controls the acceleration level to synchronize the vehicle speed with the freeway vehicles and ensure a safe gap distance [27,33]. During this lane-changing preparation process, the driver of SV repeatedly checks surroundings to judge if their decision can be fulfilled and tries to follow-up on their decision. In this study, therefore, this repetition in decision-making for merging maneuvers prior to lane-changing execution was regarded as playing the game repeatedly.

The repeated game concept implies that a stage game with identical structure is repeatedly played until termination of the game, which is divided into two classes: finite or infinite, depending on the players' beliefs about the number of repetitions. In this study, the decision-making game for merging was regarded as an infinitely repeated game because the players in the game do not know how many times the game will be repeated. Note that for an infinitely repeated game, the stage game will not necessarily be repeated an infinite number of times.

Drivers (i.e., players) interact by playing a stage game numerous times. As a summary of explanation about the game model type, the one-shot game model implies that previous game results do not affect the present game, while the decision-makers take previous game results into account in the repeated game model, as illustrated in Figure 3. This study adopts the repeated decision-making game approach using the cumulative payoffs to prevent repeated fluctuations in payoffs, as proposed in [2]. The stage decision-making game is conducted periodically and repeatedly over discrete time periods $T \in [t_1, t_n]$. Time preference is considered by assuming that future payoffs are weighted proportionately at a rate $\delta$, called the rate factor. Cumulative payoffs of the driver $d$ for action strategy $a_{ij}$, i.e., $U_{ij}^d = P_{ij}$ or $Q_{ij}$, are presented in Equation (1).

$$U_{ij}^d(T) = \sum_{t_1}^{t_n} \delta^{t-1} u_{ij}^d(t). \tag{1}$$

Here $u_{ij}^d(t)$ is a utility of a driver $d$ for an action strategy set $(s_i, l_j)$ at time step $t$; $T$ is the number of decision-making time steps; and $d$ denotes a driver, i.e., player in a game, the driver of SV or DL. If $\delta > 1$, it implies that the current payoffs are more important than the past payoffs. Otherwise, the previous game results could significantly affect the decision-making in a future game.

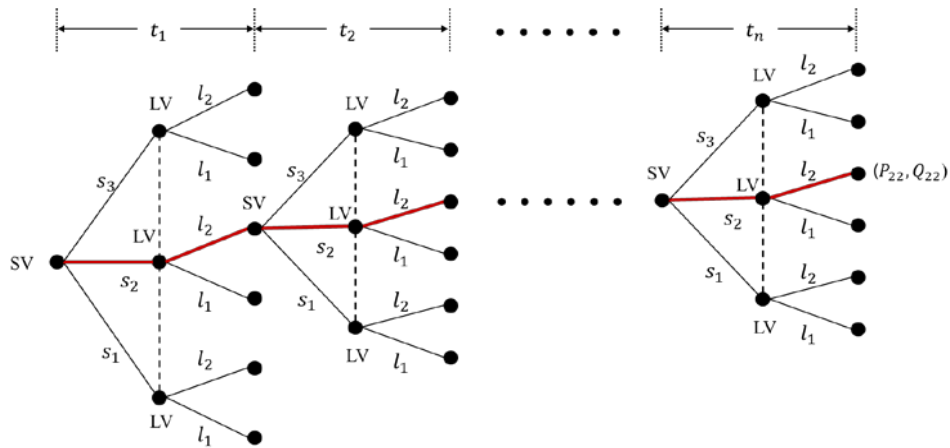

**Figure 3.** Decision-making game based on the repeated game approach in extensive form.

### 3.3. Reformulated payoff functions

In previous game theory-based-models, the payoff functions for two players were formulated using the significant decision factors, such as safety, spacing (or gap), relative speed, travel time, expected acceleration level, remaining distance to reach the end of acceleration lane, and so on [1-2,11,25,29,31]. In [1], we initially proposed the payoffs using five decision factors: minimization of travel time, avoidance of collisions (i.e., safety), travel efficiency, the LV's expected acceleration, and the remaining distance to execute the maneuver. In a following study [2], the payoffs of the driver of SV were formulated as the expected gap and remaining distance, and the expected relative speed was considered as the other driver's main decision variable. Both previous studies used multiple dimensioned variables making the payoffs as only interpreted as qualitative outcome to represent the player's preference. In addition, an error term was considered to capture unobserved variables assumed to be a constant, resulting in minimal consideration of a driver's randomness. As described previously, therefore, this study updates the payoff functions to use efficient decision variables including a random error term and proposes monotone (dimensionless) functions by transformation of quantitative variables. This section introduces the decision variables, and then presents the reformulated payoff functions for each driver.

#### 3.3.1. Safety payoff

Among various decision factors, safety is a key factor for human drivers' decisions to avoid a potential collision or not induce a dangerous situation. Yu et al. [31] used the time headway as a safety payoff, as presented in Equation (2).

$$h_{PV,SV}(t) = \frac{x_{PV}(t) - x_{SV}(t)}{v_{SV}(t)}, \tag{2}$$

Here $x_{PV}(t)$ and $x_{SV}(t)$ are the positions of the (potential) PV and SV at instant time $t$, respectively; and $v_{SV}(t)$ is speed of the SV at instant time $t$. However, they did not take the speed of a PV into account. In [2], the expected spacing between vehicles, indicating a possibility to ensure safe distance with consideration of vehicles' speed and acceleration levels, was used. Additionally, Wang et al. [34] used a penalty formulated using relative speed and the gap distance. Kita [25] used the TTC between vehicles as the main payoff, as defined in Equation (3).

$$TTC_{PV,SV}(t) = \frac{x_{PV}(t) - x_{SV}(t) - l_{PV}}{v_{SV}(t) - v_{PV}(t)} \qquad \text{if} \quad v_{SV}(t) > v_{PV}(t), \qquad (3)$$

Here $l_{PV}$ denotes the length of the PV; and $v_{PV}(t)$ is the speed of the PV at instant time $t$.

The interactive effects of relative speed and gap distance are contained in the single measure TTC [35]. Brackstone et al. [36] collected realistic data using an instrumented vehicle equipped with relative distance- and speed-measuring sensors. Observations of vehicle trajectories from five participants showed that TTC is a major factor in lane-changing decisions. Most collision avoidance systems (or pre-crash safety systems) applied in a vehicle use the instantaneous TTC to evaluate collision risk [37]. Moreover, Vogel [38] recommended the use of TTC for evaluation of safety because it indicates the actual occurrences of dangerous situations. Vogel also noted that a situation with a small TTC is imminently dangerous, and that a situation with a small headway and relatively large TTC is a potentially dangerous situation. Therefore, this study proposes the integrated safety payoff function $A^S$ with consideration of not only TTC but also headway, which was formulated using the hyperbolic tangent function, as presented in Equations (4) and (5).

$$A_{PV,SV}^S = \begin{cases} \left( tanh\left(\frac{TTC_{PV,SV}(t)}{t^S} - 1\right) + tanh\left(\frac{h_{PV,SV}(t)}{t^S} - 1\right) \right) \times 0.5, & \text{if } v_{SV}(t) > v_{PV}(t) \\ \left( 1 + tanh\left(\frac{h_{PV,SV}(t)}{t^S} - 1\right) \right) \times 0.5, & o.w. \end{cases}, \qquad (4)$$

$$A_{SV,LV}^S = \begin{cases} \left( tanh\left(\frac{TTC_{SV,LV}(t)}{t^S} - 1\right) + tanh\left(\frac{h_{SV,LV}(t)}{t^S} - 1\right) \right) \times 0.5, & \text{if } v_{LV}(t) > v_{SV}(t) \\ \left( 1 + tanh\left(\frac{h_{SV,LV}(t)}{t^S} - 1\right) \right) \times 0.5, & o.w. \end{cases}. \qquad (5)$$

Here $t^S = \min\left(\frac{RD_{SV}}{v_{SV}(t)}, 3\right)$ denotes the minimum safe time headway between the 3-second rule recommended by the National Safety Council [39] and the time headway to reach the end of the acceleration lane.

The safety payoffs of both drivers for the action strategies were formulated to satisfy $U^S \in [-1,1]$, as shown in Equations (6) to (9).

$$U_{SV}^S(s_1) = 0.5\left(A_{PV,SV}^S + A_{SV,LV}^S\right), \qquad (6)$$

$$U_{SV}^S(s_2) = -A_{SV,LV}^S, \qquad (7)$$

$$U_{SV}^S(s_3) = -A_{PV,SV}^S, \qquad (8)$$

$$U_{LV}^S(l_1) = A_{SV,LV}^S = -U_{LV}^S(l_2). \qquad (9)$$

For the 'change $(s_1)$' action of the driver of SV, $U_{SV}^S(s_1)$ was formulated as the average of safety payoffs taking both the PV and LV in the target lane into account. For the 'wait $(s_2)$' and 'overtake $(s_3)$' action of the driver of SV, on the other hand, the driver's safety payoffs were formulated to consider only the corresponding vehicle related to each action strategy. Likewise, it was assumed that the driver of LV also evaluates their safety in consideration of the SV only.

As shown in the safety payoff formulation, the safety payoffs vary by spacing between vehicles and each vehicle's speed. Figure 4 shows the prospective safety payoffs of the driver of SV at various speeds of the three vehicles (i.e., PV, SV, and LV), with the SV in different positions between the PV and LV. In this example, spacing between the PV and LV is constant at 77 m. Figure 4(a) presents the case in which the SV is located close to the PV. In other words, the lead gap $\Delta x_{PV,SV}$ is small and the lag gap $\Delta x_{SV,LV}$ is large. If $v_{PV} > v_{SV}$, $U_{SV}^S(s_1)$ is greater than $U_{SV}^S(s_3)$. Otherwise, the driver of SV is attracted to choose the 'overtake $(s_3)$' action in consideration of safety. In the second case described in Figure 4(b), the SV is located at the middle position between the PV and LV. Therefore, the 'change $(s_1)$' action is relatively attractive, i.e., $U_{SV}^S(s_1) > U_{SV}^S(s_2)$ and, $U_{SV}^S(s_1) > U_{SV}^S(s_3)$ even if $v_{SV}$ is slightly less than $v_{PV}$ and $v_{LV}$. The the 'overtake $(s_3)$' action is attractive when $v_{SV} \gg v_{PV}$, and $U_{SV}^S(s_2)$ is greater than $U_{SV}^S(s_1)$ when $v_{SV} \ll v_{LV}$. The last case, in which the SV is close to the LV, represents the case where the driver of SV is drawn to choosing the 'wait $(s_2)$' action if $v_{LV} > v_{SV}$. If $v_{SV} > v_{LV}$, the 'change $(s_1)$' action is more attractive. From these cases, transformed safety payoffs are reasonable to represent the general decision-making results of the driver of SV.

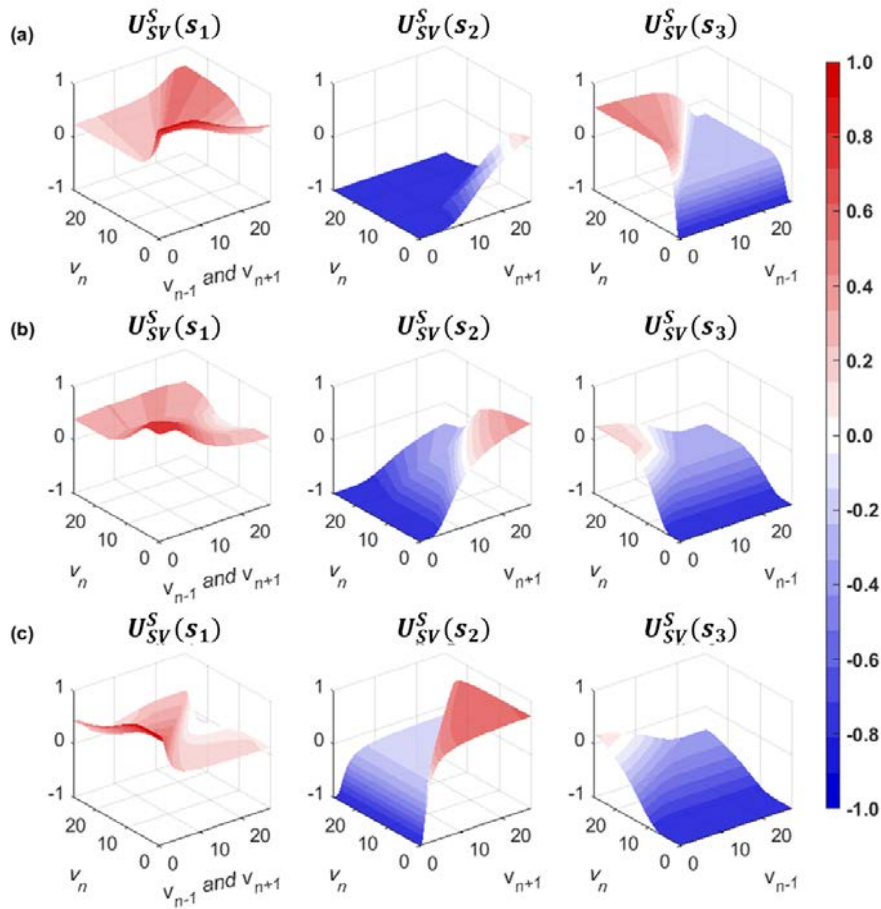

**Figure 4.** Safety payoffs of the driver of SV for the $s_1$, $s_2$, and $s_3$ action: (a) close to the PV ($\Delta x_{SV,LV} = 67\ m, \Delta x_{PV,SV} = 10\ m$); (b) middle position between PV and LV ($\Delta x_{SV,LV} = 38\ m, \Delta x_{PV,SV} = 39\ m$); (c) close to the LV ($\Delta x_{SV,LV} = 10\ m, \Delta x_{PV,SV} = 67\ m$).

Figure 5 presents the safety payoffs for the driver of LV in the three cases described above. In Figure 5(a), which shows that $\Delta x_{SV,LV}$ is considerably large, the driver of LV desires to choose the 'yield $(l_1)$' action, except in the case where $v_n \ll v_{n+1}$. These payoffs seem to be reasonable because the LV is far away from the SV. In the second case, the 'yield $(l_1)$' action is attractive as well. This case is similar to a real field situation, where the lane-changing action of the following vehicle in the target lane mostly shows cooperation in order to accept the merging vehicle's lane change. In the third case, the huge deceleration is expected to provide a gap to the SV because the LV is close to the SV. Therefore, the safety payoffs of the driver of LV for the 'block $(l_2)$' action is higher than that for the $l_1$ action if $v_{SV} < v_{LV}$. Otherwise, the safety payoff of the driver of LV for the 'yield $(l_1)$' action is slightly higher, except in a freeway congested traffic condition (i.e., $v_{SV} \gg v_{LV}$).

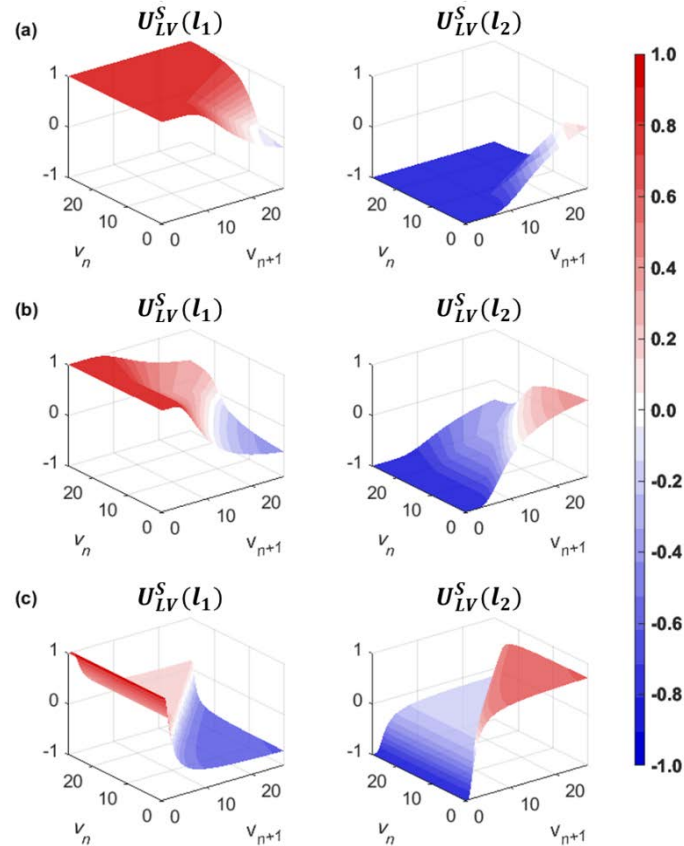

**Figure 5.** Safety payoffs of the driver of LV for the $l_1$ and $l_2$ action: (a) close to the PV ($\Delta x_{SV,LV} = 67\ m, \Delta x_{PV,SV} = 10\ m$); (b) middle position between PV and LV ($\Delta x_{SV,LV} = 38\ m, \Delta x_{PV,SV} = 39\ m$); (c) close to the LV ($\Delta x_{SV,LV} = 10\ m, \Delta x_{PV,SV} = 67\ m$).

### 3.3.2. Forced merging payoff for the driver of SV

According to the empirical field data collected at a freeway merging section, the driver of a vehicle entering through an on-ramp usually accelerates for speed-harmonization with freeway vehicles. The driver of SV then selects a gap to merge onto the freeway. In congested traffic conditions, however, the merging vehicles travel at a higher speed than the surrounding vehicles on the freeway. Thus, the driver occasionally rejects the initial gap and then uses a farther forward gap close to the end of the acceleration lane. Wan et al. found that merging vehicles pass freeway vehicles and try to find an acceptable gap to merge onto the freeway after traveling longer than the normal merging cases in congested traffic conditions [27]. Marczak et al. [40] analyzed data collected at two sites to find variables related to gap acceptance, concluding that the distance to the end of the acceleration lane is a significant variable. Also, Hwang and Park [41] concluded that the remaining distance is the most important factor for determining gap acceptance; the driver will most likely accept a smaller gap if the remaining distance to the end of the acceleration lane is smaller. In order to consider the case in which a vehicle merges close to the end of the acceleration lane, the payoff function of the driver of SV should include a term called the forced merging payoff, which relates the remaining distance to the end of the acceleration lane. It affects that the driver decides the 'change $(s_1)$' action at the decision point where the remaining distance is considerably short.

This study formulated the forced merging payoff as a function of the remaining distance and $v_{SV}(t)$. There is an assumption that the end of the acceleration lane is an imaginary preceding vehicle that is stopped. The presence of this imaginary vehicle, which is also considered as a hard wall, means the driver of SV cannot drive further due to the restricted length of the acceleration lane. Thus, the expected safety distance to maintain the instant speed of the SV, $v_{SV}(t)$, was estimated by a car-following model. This study used the RPA car-following model, which was first developed by Rakha et al. [42]. Performance of the RPA car-following model has been validated against naturalistic driving data [43]. This study estimated the safety distance for the SV, $x_{SV}^{CF}(t)$ using the RPA model's two components: steady-state traffic stream behavior and collision avoidance. The steady state modeling applies the Van Aerde's steady state car-following model [44-45], which is a non-linear single regime function of vehicle speed and spacing. The first safe spacing (i.e., safety distance) provided by the steady-state model is

$$x_{SV}^{CF_1}(t) = c_1 + c_3 \cdot v_{SV}(t) + \frac{c_2}{v_f - v_{SV}(t)}. \tag{10}$$

Here $v_f$ indicates the free-flow speed. The model coefficients can be computed as

$$c_1 = \frac{v_f}{k_j v_c^2}(2v_c - v_f), \tag{11}$$

$$c_2 = \frac{v_f}{k_j v_c^2}(v_f - v_c)^2, \tag{12}$$

$$c_3 = \frac{1}{q_c} - \frac{v_f}{k_j v_c^2}. \tag{13}$$

Here $k_j$, $v_c$, and $q_c$ indicate the jam density, speed-at-capacity, and saturation flow rate, respectively. The detailed definition of these coefficients is described in [44].

As the second component of the RPA model, collision avoidance was modeled to avoid incidents at non-steady state conditions [43]. The second safe spacing estimated by collision avoidance is defined as

$$x_{SV}^{CF\_2}(t) = \frac{v_{SV}(t)^2}{2 \cdot a_{min}} + x_j. \tag{14}$$

Here $a_{min}$ and $x_j$ denote the minimum acceleration (i.e., maximum deceleration) and the jam spacing, respectively.

The maximum value of two safe spacings, $x_{SV}^{CF\_1}(t)$ and $x_{SV}^{CF\_2}(t)$, is considered as the expected safe spacing to maintain current speed.

$$x_{SV}^{CF}(t) = \max(x_n^{CF\_1}(t), x_n^{CF\_2}(t), x_{max}^{RD}). \tag{15}$$

Here $x_{max}^{RD}$ is the maximum of the remaining distance, i.e., the longitudinal length of the acceleration lane.

To balance each payoff, this study re-formulated the forced merging payoff of the driver of SV, $U_{SV}^{FM}$.

$$U_{SV}^{FM} = \left[ \frac{\max(x_{SV}^{CF}(t) - x_{SV}^{RD}(t), 0)}{x_{SV}^{CF}(t)} \right]^2. \tag{16}$$

Here $x_{SV}^{RD}(t)$ indicates the remaining distance for the SV in the acceleration lane at time $t$. This formulation satisfies $U_{SV}^{FM} \in [0,1]$ as shown in Figure 6. If the remaining distance is shorter than $x_{SV}^{CF}(t)$, $U_{SV}^{FM}$ begins to have positive payoffs, inducing a preference for the 'change $(s_1)$' action. This term presents greater payoffs when $v_{SV}(t)$ is faster.

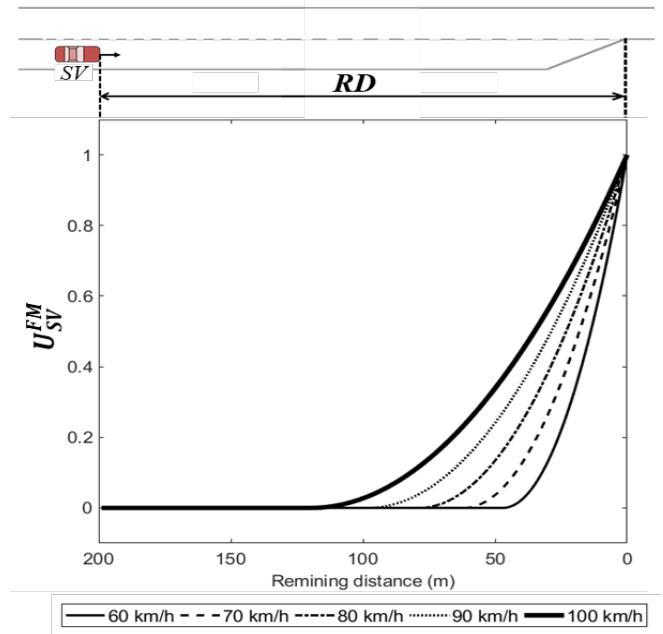

**Figure 6.** Forced merging payoff by the remaining distance at various speeds.

### 3.3.3. Payoff functions for the drivers of the SV and LV

Table 1 represents the updated merging decision-making model in normal form. The payoff functions of the driver of SV consist of both the safety and forced merging payoffs, and those of the driver of LV include the safety payoffs only. In order to capture unobserved utility, both players' payoff functions also have an error term, which was assumed to be normally distributed as $\varepsilon_{ij}^{SV\ or\ LV} \sim N(0,1)$. The parameters in the payoff functions, i.e., set of $\alpha_{ij}$ and $\beta_{ij}$ ($i$ = 1,2,3 and $j$ = 1, 2), are parameters to be estimated.

**Table 1.** Game Structure and Payoff Functions of the Merging Decision-Making Game in Normal Form

| Player & Actions | | Driver of LV | |
|---|---|---|---|
| | | **Yield [$l_1(q_1)$]** [2] | **Block [$l_2(q_2)$]** |
| **Driver of SV** | Change [$s_1(p_1)$] [1] | $P_{11} = \alpha_{11}^1 + \alpha_{11}^2 U_{SV}^S(s_1) + \alpha_{11}^3 U_{SV}^{FM} + \varepsilon_{11}^{SV}$ $Q_{11} = \beta_{11}^1 + \beta_{11}^2 U_{LV}^S(l_1) + \varepsilon_{11}^{LV}$ | $P_{12} = \alpha_{12}^1 + \alpha_{12}^2 U_{SV}^S(s_1) + \alpha_{12}^3 U_{SV}^{FM} + \varepsilon_{12}^{SV}$ $Q_{12} = \beta_{12}^1 + \beta_{12}^2 U_{LV}^S(l_2) + \varepsilon_{12}^{LV}$ |
| | Wait [$s_2(p_2)$] | $P_{21} = \alpha_{21}^1 + \alpha_{21}^2 U_{SV}^S(s_2) + \varepsilon_{21}^{SV}$ $Q_{21} = \beta_{21}^1 + \beta_{21}^2 U_{LV}^S(l_1) + \varepsilon_{21}^{LV}$ | $P_{22} = \alpha_{22}^1 + \alpha_{22}^2 U_{SV}^S(s_2) + \varepsilon_{22}^{SV}$ $Q_{22} = \beta_{22}^1 + \beta_{22}^2 U_{LV}^S(l_2) + \varepsilon_{22}^{LV}$ |
| | Overtake [$s_3(p_3)$] | $P_{31} = \alpha_{31}^1 + \alpha_{31}^3 U_{SV}^S(s_3) + \varepsilon_{31}^{SV}$ $Q_{31} = \beta_{31}^1 + \beta_{31}^2 U_{LV}^S(l_1) + \varepsilon_{31}^{LV}$ | $P_{32} = \alpha_{32}^1 + \alpha_{32}^2 U_{SV}^S(s_3) + \varepsilon_{32}^{SV}$ $Q_{32} = \beta_{32}^1 + \beta_{32}^2 U_{LV}^S(l_2) + \varepsilon_{32}^{LV}$ |

[1] $p_i$ in parentheses denotes the probability assigned to the pure strategy of the driver of SV, $s_i$; $\sum_{i=1}^3 p_i = 1$.

[2] $q_j$ in parentheses denotes the probability assigned to the pure strategy of the driver of LV, $l_j$; $\sum_{j=1}^2 q_j = 1$.

## 4. Model Calibration and Validation

Model evaluation was conducted to prove efficiency of the game models using the stage game based on the newly formulated payoff functions. This section introduces the observation dataset for model evaluation and calibration methodology. In addition, calibration and validation results of our previous model and the updated repeated game models are presented.

### 4.1. Preparation of Observation Dataset

This study used NGSIM vehicle trajectory data from a segment of U.S. Highway 101 (Hollywood Freeway) in Los Angeles, California, collected between 7:50 a.m. and 8:35 a.m. on June 15, 2005 [14-15]. Reasonable classification of the action strategies chosen by the drivers of the SV and LV is a critical issue, as it is directly related to the results of the game model [2]. There is a limitation on the classification of drivers' decisions based on trajectories and speed profile data. This study used a total of 1,504 observations extracted from NGSIM data in [2]. For classification of the SV's maneuvers observed in the field, this study used the type of gaps that were selected at game playing moments among the three following gap types (as illustrated in Fig. 1(a)): (1) forward (lead) gap, (2) adjacent (current) gap, or (3) backward (lag) gap. In addition, the spacing between the SV and LV was used for classification of the LV's maneuvers. Detailed classification methodology is described in [2]. Next, all data were reviewed to judge whether the classification results were reasonable to show drivers' intentions. If the specific data were regarded as improper classification, these data were modified. Decisions made by drivers in all observations were classified using this process.

### 4.2. Model Calibration

#### 4.2.1. Calibration approach

In the game model, each player chooses an action to achieve the goal of the game. In game theory, the Nash equilibrium is a solution to find the optimal set of strategies for both drivers where they have no incentive to deviate from their initial strategy. If the Nash equilibrium exists, it implies that each player will choose the strategy that maximizes their own payoff while considering an opponent who also wants to maximize their payoff. The Nash equilibrium defines pure strategy as

$$\begin{cases} P(s^*, \ l^*) \geq P(s_i, l^*), & \forall \ s_i \in S, i = 1,2,3 \\ Q(s^*, l^*) \geq Q(s^*, l_j), & \forall \ l_j \in L, j = 1,2 \end{cases}, \tag{17}$$

where $s^*$ and $l^*$ indicate the equilibrium action strategy of the drivers of the SV and LV, respectively. In this study, if a pure strategy Nash equilibrium does not exist, a mixed strategy Nash equilibrium involves at least one player playing a randomized strategy and no player being able to increase their expected payoff by playing an alternate strategy. A probability for each player's strategy is assigned with consideration of each player's expected payoff from the different strategies [28]. This paper used the MATLAB function NPG developed by Chatterjee [46] to solve a two-player, finite, non-cooperative game. Chatterjee's algorithm [46] solves the game by computing the Nash equilibrium in mixed strategies based on the estimated parameters and expected payoffs (i.e., $P_{ij}$ and $Q_{ij}$). The algorithm provides the probabilities of the choice of pure action strategies for each driver (i.e., $p_i$ and $q_j$) in each observation.

In order to calibrate the merging decision-making model, this study followed the calibration method developed by Liu et al. [11], who proposed a parameter estimation method by solving a bi-level programming problem. As illustrated in Figure 7, the lower-level programming is to find the Nash equilibrium using Chatterjee's function [46]. This study also defined that the upper level is a non-linear programming problem to minimize the total deviation of probabilities in the system in order to choose actual observed actions by the following function

$$\min \sum_{k=1}^{n} (1 - p_{a^k} \cdot q_{a^k}), \tag{18}$$

where $k$ denotes the index of observations; $a^k$ is the observed action strategies set $(s_i{}^k, l_j{}^k)$ in observation $k$; and $p_{a^k}$ and $q_{a^k}$ are the probabilities of the drivers of the SV and LV, respectively, choosing the observed action in $a_k$. Here, $A^k$ and $B^k$ denote all parameters to be estimated for each driver's payoff functions.

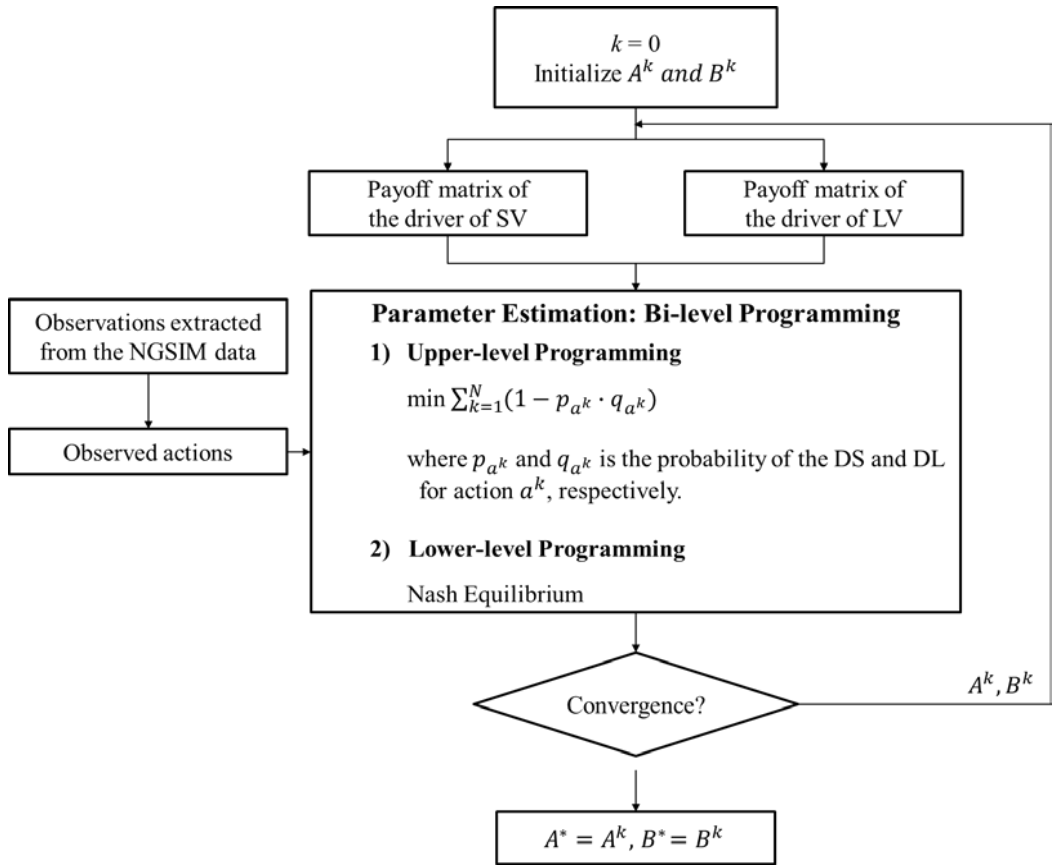

**Figure 7.** Schematic workflow for bi-level programming.

### 4.2.2 Calibration results

As mentioned earlier, this study calibrated a total of two types of game models: (1) the one-shot game model, in which the developed stage game is played independently at every game point based on instantaneous status only; (2) the repeated game model using the cumulative payoffs with factor $\delta$ of various rates. To verify performance of the updated payoff functions in predicting human-drivers' decisions in merging situations, the first type of the models was subdivided into two models according to the payoff functions used in model calibration as below.

- One-shot game model based on the stage game using the payoff functions developed in [2]
- One-shot game model based on the stage game using the reformulated payoff functions in Section 3.3

Herein the former and latter models were called as the 'previous one-shot game model' and the 'one-shot game model', respectively. For model calibration, a NGSIM dataset observed between 7:50 a.m. and 8:20 a.m. was used. The number of observations used in model calibration was 685 (out of 1,504). Table 2 shows the estimated parameters of the payoff functions of the drivers of the SV and LV.

In order to compare the models' prediction accuracy, the mean absolute error (MAE) was calculated using Equation (19).

$$MAE = \frac{1}{N} \sum_{k=1}^{N} |1 - 1(\hat{x}_k - x_k)|, \tag{19}$$

where $N$, $\hat{x}_k$, and $x_k$ denote the number of observations, model prediction, and actual observations, respectively. Note that $1(\hat{x}_k - x_k)$ is equal to one if $\hat{x}_k = x_k$, and is zero otherwise. The model prediction $\hat{x}_k$ was estimated by probabilities calculated using Chatterjee's algorithm [46]. Table 3 shows the calibration results for the MAEs of the three types of models. In comparison with our previous model, the one-shot game model using the updated payoff functions shows higher prediction capacity in merging decision-making. In the repeated game models, the models with $\delta > 1.0$ were calibrated with lower MAEs than those with $\delta \leq 1.0$.

**Table 2.** Estimated Parameters of the Payoff Functions for Game Models

| Payoff Function | Parameters | One-shot Game Model | Repeated Game Models | | | | | |
|---|---|---|---|---|---|---|---|---|
| | | | Model 1 ($\delta$=0.6) | Model 2 ($\delta$=0.8) | Model 3 ($\delta$=1.0) | Model 4 ($\delta$=1.2) | Model 5 ($\delta$=1.4) | Model 6 ($\delta$=1.6) |
| $P_{11}$ | $\alpha_{11}^1$ | 9.64 | 5.10 | 2.88 | 6.69 | -1.77 | 7.08 | 7.11 |
| | $\alpha_{11}^2$ | 23.51 | 74.83 | 48.38 | 96.45 | 9.20 | 27.34 | 8.38 |
| | $\alpha_{11}^3$ | 32.69 | 59.51 | 69.45 | 1.00 | 5.16 | 97.08 | 2.75 |
| $P_{12}$ | $\alpha_{12}^1$ | 9.43 | 8.83 | 3.58 | 7.87 | 8.64 | 7.27 | -6.26 |
| | $\alpha_{12}^2$ | 87.57 | 77.60 | 44.40 | 86.30 | 3.11 | 50.13 | 4.25 |
| | $\alpha_{12}^3$ | 10.98 | 43.84 | 1.80 | 71.19 | 5.73 | 84.75 | 7.34 |
| $P_{21}$ | $\alpha_{21}^1$ | 0.63 | -9.78 | -7.49 | -6.91 | -8.88 | -6.65 | -8.13 |
| | $\alpha_{21}^2$ | 3.35 | 26.60 | 10.68 | 62.49 | 3.18 | 31.94 | 1.75 |
| $P_{22}$ | $\alpha_{22}^1$ | -7.88 | -8.50 | -3.42 | -6.19 | 9.73 | -8.98 | 5.56 |
| | $\alpha_{22}^2$ | 42.64 | 20.75 | 5.21 | 65.72 | 6.22 | 19.43 | 7.16 |
| $P_{31}$ | $\alpha_{31}^1$ | -0.66 | 6.07 | -9.38 | -6.21 | -2.84 | -5.18 | 6.41 |
| | $\alpha_{31}^2$ | 67.24 | 48.05 | 78.92 | 94.59 | 11.19 | 25.08 | 7.53 |
| $P_{32}$ | $\alpha_{32}^1$ | -0.53 | -3.10 | -5.39 | -0.44 | 2.75 | -3.69 | 8.35 |
| | $\alpha_{32}^2$ | 16.91 | 52.79 | 95.22 | 59.86 | 2.21 | 30.06 | 4.79 |
| $Q_{11}$ | $\beta_{11}^1$ | 9.93 | 3.78 | 6.96 | 9.80 | -1.99 | 7.97 | -3.75 |
| | $\beta_{11}^2$ | 13.30 | 17.29 | 6.64 | 25.06 | 6.88 | 5.86 | 10.22 |
| $Q_{12}$ | $\beta_{12}^1$ | -1.26 | -8.39 | -6.24 | -5.83 | -7.03 | -8.90 | -8.36 |
| | $\beta_{12}^2$ | 3.70 | 0.29 | 19.40 | 23.84 | 10.20 | 18.49 | 1.89 |
| $Q_{21}$ | $\beta_{21}^1$ | 5.78 | 7.64 | 8.05 | 8.74 | 5.52 | 8.25 | 0.27 |
| | $\beta_{21}^2$ | 89.18 | 57.76 | 58.65 | 78.06 | 2.76 | 82.45 | 4.12 |
| $Q_{22}$ | $\beta_{22}^1$ | 7.73 | -4.36 | -4.36 | 0.63 | 0.34 | -8.66 | -5.95 |
| | $\beta_{22}^2$ | 57.97 | 6.64 | 55.26 | 14.12 | 7.43 | 38.74 | 7.61 |
| $Q_{31}$ | $\beta_{31}^1$ | 3.88 | -4.02 | -6.99 | 6.38 | 9.39 | -0.82 | 3.68 |
| | $\beta_{31}^2$ | 55.87 | 96.95 | 98.01 | 1.12 | 4.35 | 46.49 | 9.22 |
| $Q_{32}$ | $\beta_{32}^1$ | 4.26 | -9.75 | 1.08 | -8.01 | 6.78 | 1.53 | -4.85 |
| | $\beta_{32}^2$ | 27.87 | 26.74 | 22.93 | 74.89 | 2.20 | 86.19 | 7.83 |

Note that the previous one-shot game model using the payoff functions in [2] was calibrated using the same calibration methodology, but the estimated parameters are not shown in the table because of the different formulation for payoff functions.

**Table 3.** Calibration Results

| Models | Previous One-shot Game Model (2018) | One-shot Game Model | Repeated Game Models | | | | | |
|---|---|---|---|---|---|---|---|---|
| | | | Model 1 | Model 2 | Model 3 | Model 4 | Model 5 | Model 6 |
| Rate factor, $\delta$ | *na* [1] | *na* | 0.6 | 0.8 | 1.0 | 1.2 | 1.4 | 1.6 |
| MAE [2] | 0.2555 (74.45 %) | 0.1241 (87.59 %) | 0.1708 (82.92 %) | 0.1606 (83.94 %) | 0.1606 (83.94 %) | 0.1372 (86.28 %) | 0.1358 (86.42 %) | 0.1460 (85.40 %) |

[1] Not applicable.

[2] The number in parentheses indicates prediction accuracy.

### 4.3. Model Validation

The rest of the data, 819 observations out of 1,504, collected between 8:20 a.m. and 8:35 a.m., were used for validating the model, and the validation results are shown in Table 4. Model validation results, which show the same trends as the calibration results, are summarized as follows. First, when comparing the results of the stage game developed in the previous study [2] and this study, the prediction accuracy increase by about 12% when the third stage game is used. Thus, this study enhances the decision-making game model's performance by using the reformulated payoff functions to represent merging maneuvers. Next, in the validation results, the repeated game models with $\delta \geq 1.0$ show prediction accuracy of higher than 85%. In particular, the repeated game model shows the highest prediction accuracy when $\delta = 1.4$. Both the one-shot game and repeated game model with $\delta = 1.4$ show considerably high prediction accuracy of more than 86%. Due to limitations of unbalanced observation data [1], nevertheless, model validation using field data cannot provide evidence to be beneficial using the repeated game. It is also hard to show the apparent difference between the one-shot game and the repeated game model. In the following sections, therefore, the game models are additionally evaluated through sensitivity analysis and simulation study.

**Table 4.** Validation results

| Models | Previous One-shot Game Model (2018) | One-shot Game Model | Repeated Game Models | | | | | |
|---|---|---|---|---|---|---|---|---|
| | | | Model 1 | Model 2 | Model 3 | Model 4 | Model 5 | Model 6 |
| Rate factor, $\delta$ | *na* | *na* | 0.6 | 0.8 | 1.0 | 1.2 | 1.4 | 1.6 |
| MAE [1] | 0.2418 (75.82 %) | 0.1197 (88.03 %) | 0.1954 (80.46 %) | 0.1758 (82.42 %) | 0.1465 (85.35 %) | 0.1368 (86.32 %) | 0.1307 (86.94 %) | 0.1355 (86.45 %) |

[1] The number in parentheses indicates prediction accuracy.

## 5. Sensitivity Analysis of the Calibrated Stage Game

In this section, this study describes the sensitivity analysis conducted to observe how factor changes related to the proposed payoffs impact the stage game results. In reality, drivers' merging behavior to select an acceptable gap size and speed difference between the freeway mainline vehicles and the merging vehicle is different depending on the merging point [27,40]. Hence, this sensitivity analysis is required to demonstrate whether the developed stage game model represents merging behaviors observed in the field in various conditions. To show the decision-making model's sensitivity, the stage game is independently played in diverse scenarios varied by three input factors: game location, relative speed, and spacing. Preparation for the sensitivity analysis is presented first in the following sections, then results and corresponding discussions are provided.

*5.1. Sensitivity Analysis Setting*

As shown in Figure 8, a freeway segment that included an on-ramp was used for the analysis, with locations to play a game classified into two areas: the beginning of the acceleration lane and the end of the acceleration lane. For the spacing factor test, the SV changed its position between the PV and LV. For the speed profile test, the freeway mainline vehicles' speed was basically categorized into five scenarios: 60 $km/h$, 70 $km/h$, 80 $km/h$, 90 $km/h$, and 100 $km/h$. In each speed scenario, the SV's speed varied from 60 $km/h$ to 100 $km/h$. The freeway testbed and calibrated stage game were modeled on MATLAB, and other simulation settings are described below.

1. The length of the acceleration lane was 250 $m$.
2. Based on initial longitudinal coordination, $n-1$, $n$, and $n+1$ denote the PV, SV, and LV, respectively.
3. It was assumed that spacing between the PV and LV, $\Delta x_{n-1,n+1}$, was constant as 40 $m$: In the game played at the beginning of the acceleration lane, the PV and LV were located at 70 $m$ and 30 $m$ from the beginning of the acceleration lane, respectively. In the game played at the end of the acceleration lane, the longitudinal position of the PV and LV were 230 $m$ and 190 $m$ from the beginning point, respectively.
4. The length of all vehicles was assumed as constant at 4.8 $m$.
5. Link properties for the freeway are as follows. Saturation flow rate was 2,400 $veh/h/lane$. Jam density was 160 $veh/km/lane$. Free-flow speed and speed-at-capacity were 100 $km/h$ and 80 $km/h$, respectively.
6. Calibrated parameters of payoff functions for the repeated game model with $\delta = 1.4$ were used.

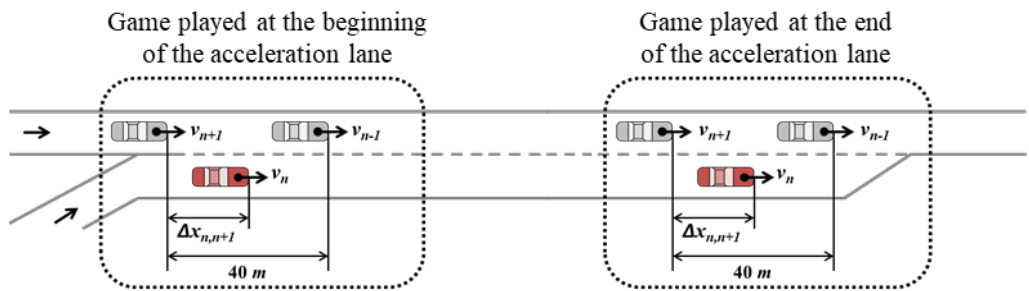

**Figure 8.** Topology of freeway merging section for sensitivity analysis.

*5.2. Sensitivity Analysis Results*

Based on the results of the stage game played at two locations in various lag spacing and relative speed scenarios, the impact of input factors and other findings revealed by the sensitivity analysis are provided. Figure 9(a) to 9(e) show the results after playing games near the beginning of the acceleration lane, and Figure 9(f) to 9(j) reveal the game results after playing the game near the end of the acceleration lane. The Chatterjee function for finding the Nash equilibrium was used to decide these game results [46]. If the game result in each case is a pure strategy Nash equilibrium, the corresponding action set is a dominant decision made by two drivers, i.e., the probability of one of six action strategies ($p_{ij} \times q_{ij}$) is one. Otherwise, when a mixed strategy Nash equilibrium exists, the game result is randomly chosen by probabilities.

Differences in drivers' behaviors based on the merging point are distinct in merging maneuver decisions. At near the beginning of the acceleration lane, a merging vehicle driver usually passes a lead vehicle when $v_n > v_{n-1}$ and when lead spacing ($\Delta x_{n-1,n}$) is quite small [27]. The higher psychological pressure related to merging makes drivers accept smaller gaps as they arrive nearer the end of the auxiliary lane compared to cases where they can take an original gap near the beginning of the acceleration lane [27]. In other words, field data show that the driver of SV tried a forced merging maneuver at close to the end of the acceleration lane [27,33]. When $v_n < v_{n+1}$ and the lag spacing ($\Delta x_{n,n+1}$) is quite small, the driver of SV waits until the LV passes the SV and then may merge using a backward gap. In Fig. 8, the calibrated stage game results show these behaviors in choosing an 'overtake ($s_3$)' and 'wait ($s_2$)' action according to the game location.

At near the beginning of the lane, as illustrated in Figure 9(a) to 9(d), the game results show that the driver of SV chooses the 'overtake ($s_3$)' action in conditions indicative of higher relative speed and short lead spacing. In contrast, the game results (as illustrated in Figure 9(f) to 9(i)) show that the driver of SV intentionally changes a lane due to a short remaining distance in the acceleration lane. For the 'wait ($s_2$)' action, differences in the results of the stage game for merging decision-making are revealed according to game location. These results prove that the forced merging utility works correctly when the SV is close to the end of the acceleration lane. Consequently, the stage game developed in this study accurately depicts realistic decisions made by human drivers according to game location.

As discussed in Section 3.3.3, TTC is critical in making lane-changing decisions. Since TTC is comprised of spacing (i.e., space headway) and relative speed, both are important in human drivers' decision-making for merging maneuvers at freeway merging sections. Hence, this study also analyzed the impacts of these factors. In Figure 9(c), blue lines parallel to the y-axis (as marked with ① to ③) and green lines parallel to the x-axis (as marked with A and B) denote test cases for sensitivity analysis on relative speed and spacing, respectively.

In the sensitivity analysis on relative speed, the PV and LV are supposed to drive at 80 $km/h$, and the SV's speed varies from 60 $km/h$ to 100 $km/h$. Scenarios were prepared with three lag spacings: 10 $m$, 20 $m$, and 30 $m$, and the game results of all scenarios are shown in Figure 10. Game results clearly show that the relative speed affects decision-making. When lag spacing ($\Delta x_{n,n+1}$) is 10 $m$ (as shown in Figure 10(a)), the drivers of the SV and LV decide on a 'wait ($s_2$) and block ($l_2$)' action set if $\Delta v_{n,n+1} \leq -10$ $km/h$. In addition, both drivers are willing to choose a 'change ($s_1$) and yield ($l_1$)' action set through the stage game if $\Delta v_{n,n+1} \geq -7$ $km/h$. These cooperative action strategy sets are results of both drivers' common consent subject to safety. In a certain range, i.e., $-10$ $km/h <$ $\Delta v_{n,n+1} < -7$ $km/h$, drivers' desired actions are competitive; in these conditions, the non-cooperative behaviors, 'change ($s_1$) and a block ($l_2$)' action, will be carried out.

When $\Delta x_{n,n+1} = 20$ $m$, in Figure 10(b), the driver of the SV and LV choose a cooperative action strategy ($s_1, l_1$) even if $\Delta v_{n,n+1} = -20$ $km/h$. This means that the relative speed is largely irrelevant in influencing the driver of SV to choose a lane-changing action if there is sufficient spacing between vehicles. If there is enough space headway, real-life experience generally shows that a driver of a merging vehicle will change a lane upon reaching an acceleration lane even though a speed harmonization process is required. In response to the merging vehicle's lane change, the driver of LV decreases speed to adjust to the new preceding vehicle (i.e., the SV) or changes a lane to the left to maintain its speed. When $\Delta x_{n,n+1} = 30$ $m$ (i.e., $\Delta x_{n-1,n} = 10$ $m$), moreover, the game results show a distinct feature depending on the relative speed. The cooperative action strategy ($s_1, l_1$) is chosen by the stage game until $v_n$ is slightly higher than $v_{n-1}$. If $\Delta v_{n,n-1} \geq 8$ $km/h$, the driver of SV chooses an 'overtake ($s_3$)' action due to a relatively small TTC in order to avoid harsh braking. Of the overtaking vehicles, 97.7% were found to have a speed higher than the freeway mainline vehicles [27]. Thus, this game model can reasonably represent decision-making results according to relative speed.

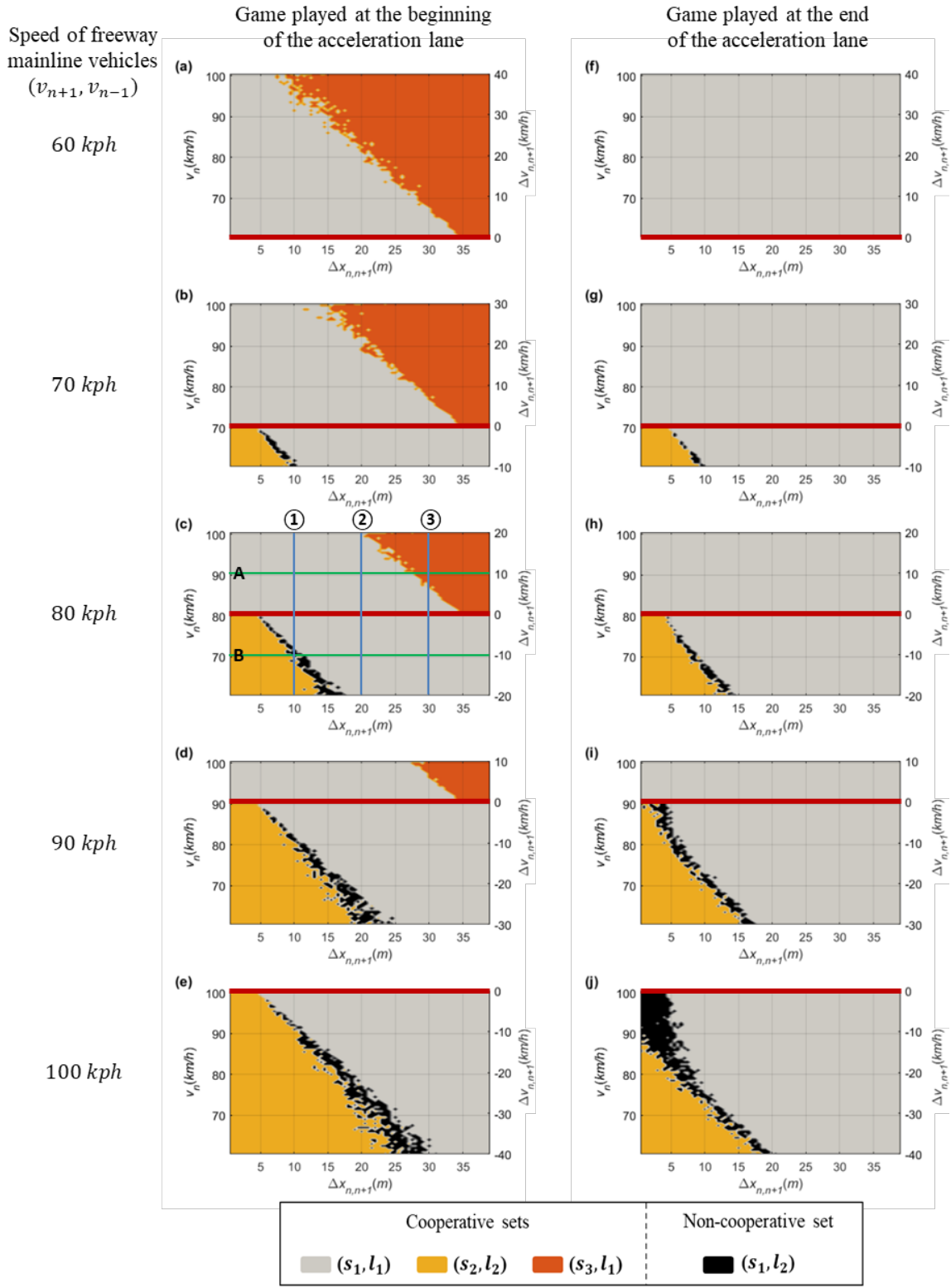

Note that a red line parallel to the x-axis on each graph indicates the speed of the freeway mainline vehicles ($v_{n-1}$, $v_{n+1}$).

**Figure 9.** Graphical representation of the one-shot game results depending on game locations, spacing between vehicles ($\Delta x_{n,n+1}$), and speed of the SV ($v_n$).

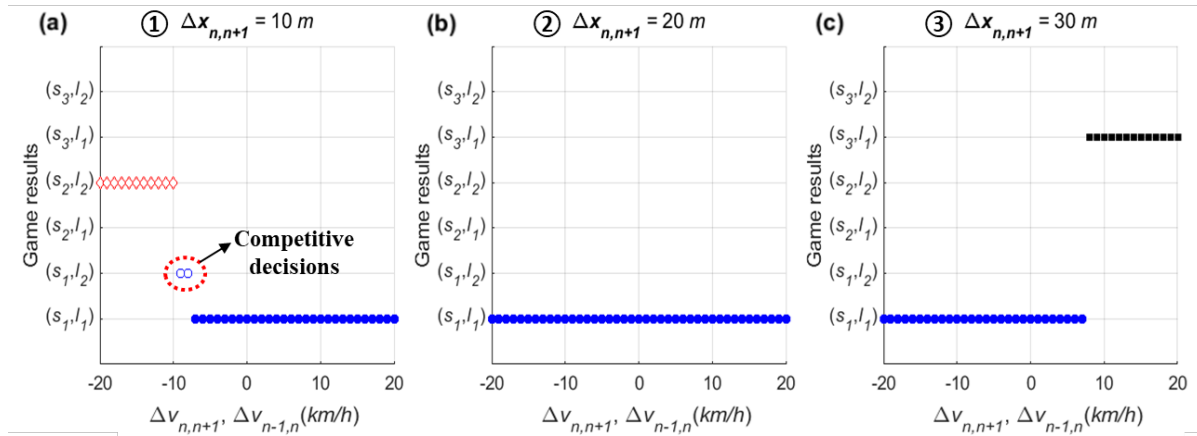

**Figure 10.** Game results on relative speed.

For the sensitivity analysis on spacing, the stage game was played with various lag spacing from $0\ m$ to $40\ m$. The PV and LV are supposed to drive at $80\ km/h$, and the SV's speed is $70\ km/h$ and $90\ km/h$. Game results of all scenarios are shown in Figure 11. In the figure, the x-axis indicates the lag spacing ($\Delta x_{n,n+1}$), and hence an increase of $\Delta x_{n,n+1}$ means a decrease of lead spacing ($\Delta x_{n-1,n}$).

When $v_n < v_{n-1}$, as shown in Figure 11(a), the stage game results show that the driver of SV decides on a 'wait ($s_2$)' action in cases in which lag spacing is less than $10\ m$. In other words, results indicate that a slower SV requires spacing of more than $10\ m$ to choose a 'change ($s_1$)' action. Depending on the spacing, competitive decision-making is also expected. This trend is also found in choosing an 'overtake ($s_3$)' action when $v_n > v_{n-1}$. In Figure 11(b), the driver of SV decides to overtake at $\Delta x_{n-1,n} \le 12\ m$. Therefore, the sensitivity results indicate that the stage game reasonably explains the difference in drivers' choices according to spacing.

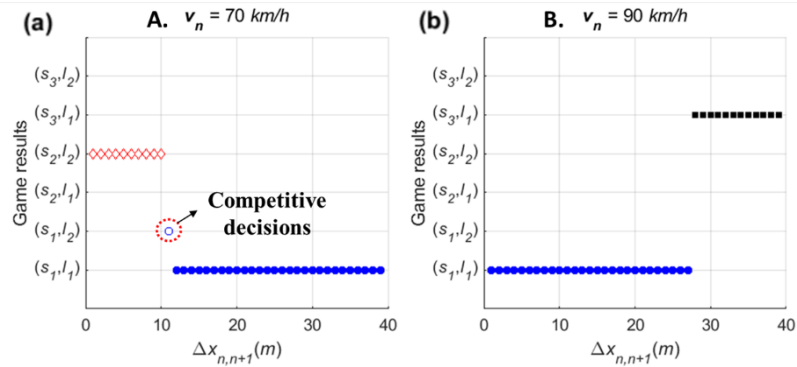

**Figure 11.** Game results on spacing.

In the results, decisions included in a non-cooperative action strategy set, i.e., $(s_1, l_2)$, are found in a specific decision-making region, as colored black in Figure 9. This region implies that this strategy set, which is decided simultaneously by drivers, puts them into competition. This result means that the driver of SV wants to change a lane after trying to ensure a safe lead and lag gap and the driver of LV does not allow the SV to merge. During the game period, one driver should change their initial decision to avoid a potential collision, and the final decision set would be a cooperative set. In addition, due to an unbalance in the number of observations indicating each action strategy, the $(s_2, l_1)$ action cannot be determined in this sensitivity analysis. From field data, including NGSIM data, it is clear that merging maneuvers are usually cooperative, as the driver of LV perceives the SV's lane-changing intention. Compared to cooperative merging, non-cooperative cases are only occasionally observed. The stage game results describe cooperative behaviors, and competition between drivers can be found at certain relative speed and spacing profiles. Consequently, the stage game model proposed in this study successfully explains rational human drivers' decision-making results.

## 6. Simulation Case Study

In this chapter, a simulation study is presented to demonstrate the performance of the game model based on the developed stage game for merging. For this case study, a microscopic simulation model based on an ABM method that included a vehicle acceleration controller was developed. To verify the performance of the ABM, a comparison between NGSIM data and simulation results is provided. The simulation setting is defined, and then various merging scenarios representing both cooperative and non-cooperative cases are explained. Next, simulation results for each scenario are presented.

### 6.1. Simulation model development

To investigate whether the repeated game model is efficient to use in microscopic traffic simulation, we used an ABM approach. ABM is a powerful method for making simulations that is widely applied across real-life problems [47-49]. This study developed a simulation model that was built on MATALB using the ABM method combined with the game model. ABM is a suitable approach for simulating the actions and interactions of intelligent entities, which includes individual people. Collaboration and competition, in particular, are major concerns in game theory; these are two typical types of human interactions addressed in several ABM methods [50]. One of the applicable situations for using ABM is when interactions among agents are heterogeneous and can lead to network effects [48,51]. Thus, this study develops a simulation model to explain merging interactions.

According to Zheng et al. [49], the ABMs explored for the existing transportation system in today's literature, in general, have the distinguishing feature of integration combining three components: drivers' action decisions, drivers' route decisions, and microsimulation. As a microsimulation component, the simulation model developed in this study basically simulates vehicle movements based on position and by speed profile as determined by an acceleration controller at each time step. As shown in Figure 12, the controller consists of a game module and a car-following module. According to the game model for the drivers' action decision component, a driver of SV plays a stage game with a driver of LV in the target lane. Depending on the action strategies at each game time, both drivers determine the acceleration level to accomplish their own strategy. In the car-following module, in addition, the desired acceleration level is decided by the RPA car-following model. In this acceleration controller, neither the individual demographic nor travel characteristics of either agent are considered.

As the game results show, when the driver of SV chooses a 'change ($s_1$)' action, they evaluate lead and lag spacing for gap acceptance to satisfy sufficient spacing and collision avoidance. If the instantaneous gap is enough to change a lane, the SV begins merging onto the freeway, and the driver of LV determines the acceleration level to follow the SV in the car-following model in response to recognition of the SV's lane-change. In addition, a route decision module is not required because merging scenarios are tested on the one-lane freeway network, which includes a merging ramp.

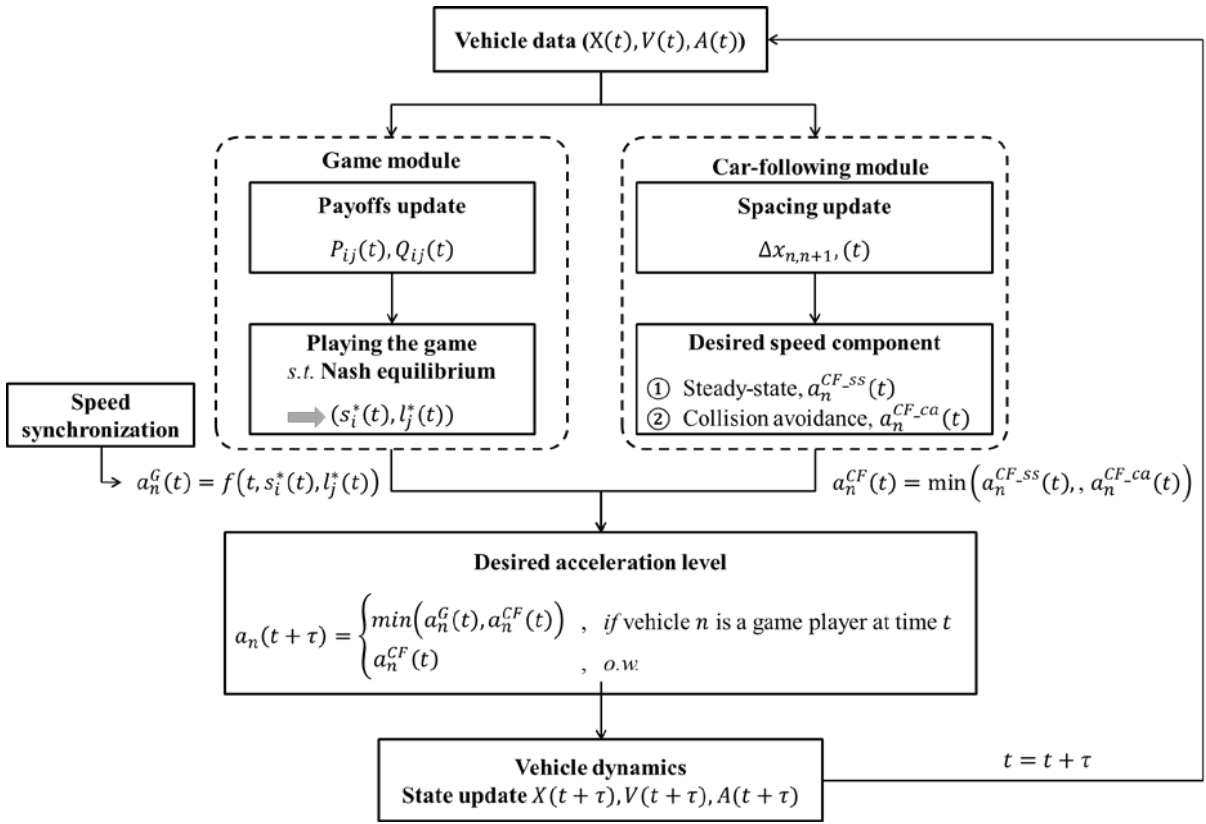

**Figure 12.** Vehicle acceleration controller structure in the developed simulation model.

Car-following module estimates an desired acceleration level based on instantaneous spacing between vehicles and speed at each time step $t$. This study used two components, i.e., steady-state and collision avoidance, of the RPA car-following model for the module [43]. The detailed definition and formulas of the components in the RPA model are described in [43]. Figure 13 shows performance of car-following module in a case which five vehicles formed a platoon. Vehicles decide an acceleration level to follow preceding vehicle by the RPA car-following model. Here, it was assumed that vehicles were located with shorter spacing than the steady-state spacing of Van Aerde's car-following model [44] at simulation time 0. As illustrated in Figure 13, therefore, following vehicles initially decreased speed to ensure proper spacing between vehicles. Then, they began to accelerate after ensuring the sufficient spacing by sequence in the platoon. In conclusion, acceleration level and speed were oscillated for a while, and then they were stabilized.

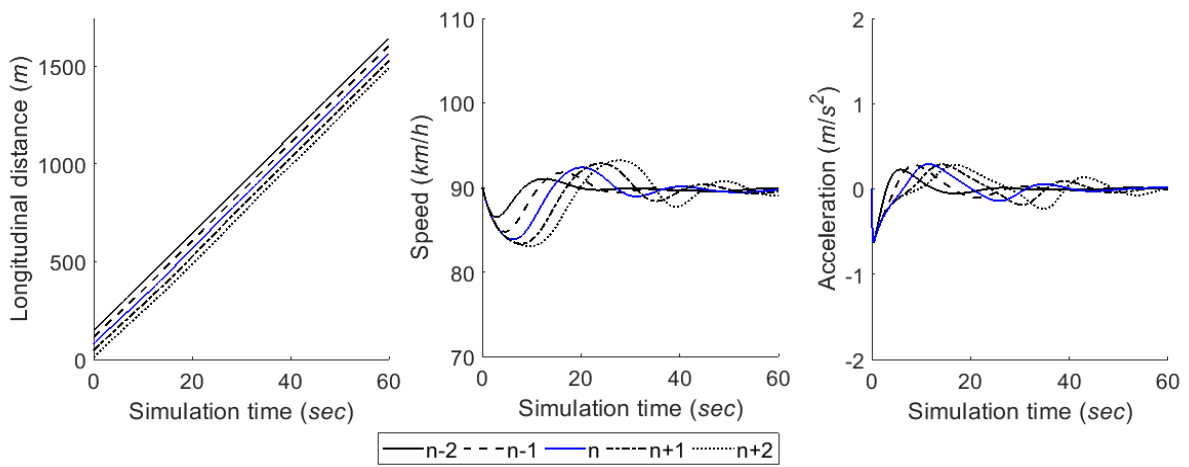

**Figure 13.** Performance of the car-following module.

The game module begins operating as soon as the SV enters the acceleration lane. The nearest following vehicle in the target lane becomes the opposite player. In this module, there are two types of merging game: (1) the one-shot game; (2) the repeated game. In detail, the one-shot game uses instantaneous payoffs, which are computed based on spacing and speed profile at time $t$, for each action strategy set, i.e., $P_{ij}(t)$, $Q_{ij}(t)$. In the repeated game, on the other hand, the cumulated payoffs are utilized. Regardless of the game type, two players decide an action strategy set subject to the Nash equilibrium. Based upon the action chosen at time $t$, the desired acceleration level for each vehicle is calculated to execute that vehicle's individual action strategy. For the SV, the desired acceleration level is determined, as stated below.

- For 'change $(s_1)$' action, the driver of SV determines acceleration level in consideration of not only speed synchronization but also gap acceptance. If $v_n(t) \ll v_{n+1}(t)$, an acceleration level for speed harmonization is additionally calculated. Also, by gap acceptance rule, another acceleration level is calculated to ensure sufficient gap for lead and lag spacing.
- For 'wait $(s_2)$' action, a required acceleration level to wait in acceleration lane until the lag vehicle pass the SV is computed. Generally, waiting cases are observed when $v_n(t) \ll v_{n+1}(t)$ and $\Delta x_{n,n+1}$ is not sufficient. If $v_n(t) \ll v_{n+1}(t)$ and the remaining distance to the end of the acceleration lane at time $t$, $RD_n(t)$, is sufficient to not require deceleration, the SV slightly accelerates to harmonize the speed with freeway vehicles during waiting time.
- Lastly, it needs to calculate the required acceleration level to use the forward gap for 'overtake $(s_3)$' action. This case is observed when $v_n(t) \gg v_{n+1}(t)$ and $\Delta x_{n-1,n}$ is not sufficient. For this strategy, therefore, speed harmonization is exclude as an acceleration component.

In addition, the driver of LV decides the acceleration level for a 'yield $(l_1)$' action by accepting the SV's merging intention. To provide safe spacing for merging, the LV's acceleration level was calculated based on the car-following model with an assumption that the SV became a potential lead vehicle. For a 'block $(l_2)$' action, on the other hand, the driver of SV shows acceleration to pass the SV by decreasing spacing. This decrease in spacing is regarded as blocking intention.

## 6.2. Simulation Model Validation

Prior to conducting a case study, validation of the simulation model developed in this study was required to determine whether the conceptual model is a reasonably accurate representation of the real world [52] and whether the output of simulations is consistent with real world output [53]. To validate the simulation model, this study used the graphical comparison technique, in which the graphs of values derived from the simulation model over time are compared with the graphs of values collected in a real system. It is a subjective, yet practical approach, and is especially useful as a preliminary approach [54]. Since the objective of the case study was to verify the repeated game's efficiency, the simulation focuses on presenting microscopic vehicle movements based on rational drivers' decision-making without consideration of individual characteristics. Considering this objective, a mathematical approach, such as statistical testing of simulation results, was not selected for model validation. Therefore, this study provides a graphical comparison between NGSIM data and the results derived from the simulation model to investigate similarity of trend in vehicle position and corresponding spacing.

This study extracted game cases from NGSIM data in which there was no interference by other surrounding vehicles except for the three main vehicles (i.e., the SV, PV, and LV). Next, instantaneous vehicles' location and speed at prior to 1.0 seconds in each case were prepared as input data for simulation. The graphical comparison results are presented showing longitudinal vehicle position and spacing are shown in Figure 14. In an example to show changing situation (see Figure 14(a)), vehicle position and corresponding lead and lag spacing are almost identical. In an example in overtaking situation (see Figure 14(b)), furthermore, considerable similarity is observed. The results show that the simulation model based on the ABM represents values similar to those found in the NGSIM data in longitudinal vehicle position and spacing. Consequently, it was possible to conclude that the developed simulation model could be utilized in the case study.

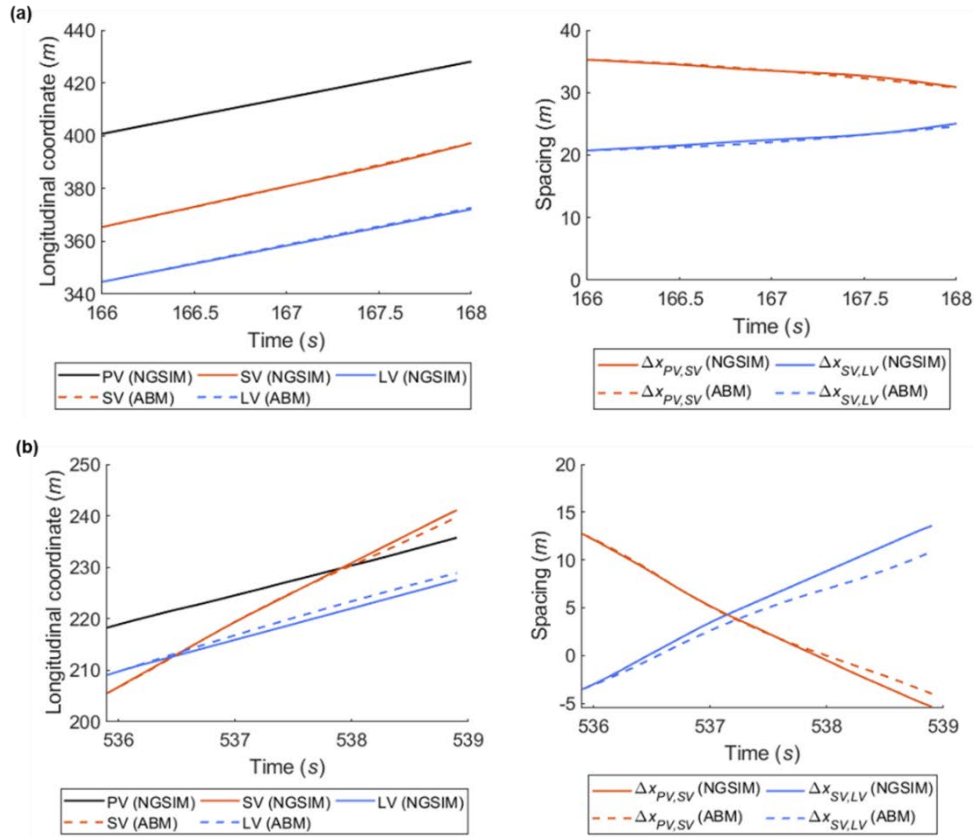

**Figure 14.** Simulation model validation results based on the graphical comparison method: (a) changing situation (SV ID: 268, PV ID: 258, and LV ID: 269 in the US101 data collected from 8:05 a.m. to 8:20 a.m.) and (b) overtaking situation (SV ID: 1108, PV ID: 1112, and LV ID: 1118 in the US101 data collected from 8:20 a.m. to 8:35 a.m.).

*6.3. Simulation setting and cases*

This study conducted case studies in various merging scenarios simulated for a total of five vehicles, including a merging vehicle. Simulation experiments were executed using both the one-shot game model and the repeated game model. As described above, the one-shot game herein is played independently without consideration of previous results at every decision-making point. The repeated game is played based on the cumulative payoffs proposed in Section 3.4. In addition, a freeway segment, including one merging section, was modeled on MATLAB, as illustrated in Figure 15. The length of the freeway mainline was 1.0 $km$ and the 250 $m$ acceleration lane was located 80 $m$ downstream of the beginning of the network. The details of the simulation settings are defined as follows.

1. Link properties for the freeway are as follows. Saturation flow rate was 2,400 $veh/h/lane$. Jam density was 160 $veh/km/lane$. Free-flow speed and speed-at-capacity were 100 $km/h$ and 80 $km/h$, respectively.

2. Based on initial longitudinal coordination, vehicles on the network were designated as $n-2$, $n-1$, $n$, $n+1$, and $n+2$, respectively. Here the vehicle $n$ denotes the SV.

3. It was assumed that the average initial speed of freeway vehicles was $v_{fwy}$. The initial speeds of four vehicles on the freeway mainline (i.e., $n-2, n-1, n+1, n+2$) were randomly determined using the normal distribution with a mean of $v_{fwy}$ and standard deviation of 0.2 at simulation start time.

4. The initial spacing between freeway vehicles, i.e., $\Delta x_{n-2,n-1}, \Delta x_{n-1,n+1}, \Delta x_{n+1,n+2}$, was determined using the Van Aerde's steady-state model according to instantaneous speed of corresponding following vehicle at time step 0.

5. With regard to the game, the time interval for playing the game was $0.5\ s$. The stage game would be newly formed if the LV or PV changed.
6. Maximum and minimum accelerations are $3.4\ m/s^2$ and $-3.4\ m/s^2$, respectively, as determined with reference to the NGSIM data. Length of all vehicles was assumed as constant as $4.8\ m$.
7. In this simulation model, the freeway mainline vehicles' behaviors to avoid a potential collision with the merging vehicle, i.e., lane change to left or deceleration before arriving the merging section, were excluded. These behaviors could not be modeled for an individual vehicle's driving maneuvers in traffic simulator because they are a result of vehicles' independent decisions rather than any interaction with the merging vehicle after recognizing the merging vehicle.

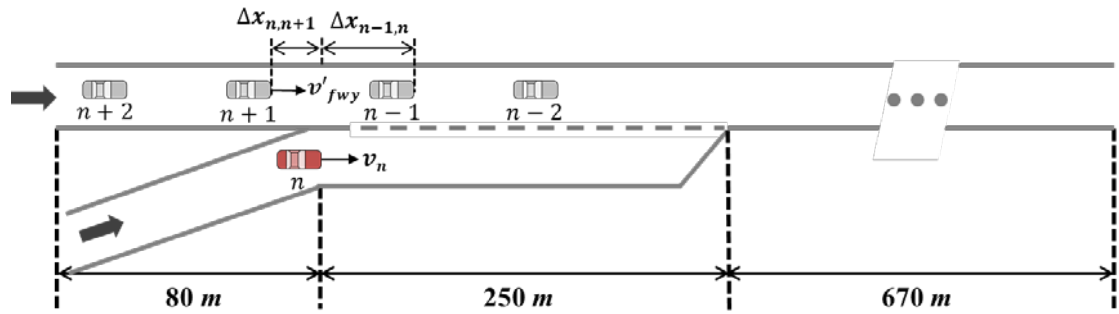

**Figure 15.** Simulation network configurations.

A total of five simulation cases were prepared, as summarized in Table 5, to represent plausible merging cases as defined by diverse input values of three factors: freeway mainline vehicles' average speed ($v_{fwy}$), initial SV's speed ($v_n$), and initial lag spacing ($\Delta x_{n,n+1}$). There are two main categories in merging: cooperative and competitive merging. Cooperative merging cases, in which the drivers' decision set would be collaborative by common consent of both drivers, indicate typical cases to select a gap type among three types: a forward gap, an adjacent gap, and a backward gap. In contrast, a competitive merging case represents an example showing a conflict in both drivers' behavior. For example, the driver of SV who wants to use an adjacent gap is willing to prepare to merge onto freeway by turning a signal on, and then executing a lane change. In that time, the driver of LV decides not to allow the cut-in to avoid the expected considerable deceleration. One of the drivers should change their initial decision in order to avoid a potential collision. This competitive situation is not common, but many drivers may have had an experience of this type. Thus, we picked two cases in order to show not only the game model's performance in non-cooperative cases but also differences between the two game models in competitive scenarios.

**Table 5.** Initial Conditions of Merging Scenarios for Case Study

| Index | Scenarios | Gap type used for merging | $\overline{v_{fwy}}$ | $\overline{v_n}$ | $\overline{\Delta x_{n,n+1}}$ |
|---|---|---|---|---|---|
| 1 | | Adjacent gap | $90\ km/h$ | $75\ km/h$ | $20.0\ m$ |
| 2 | Cooperative | Backward (lag) gap | $90\ km/h$ | $65\ km/h$ | $15.0\ m$ |
| 3 | | Forward (lead) gap | $50\ km/h$ | $65\ km/h$ | $15.0\ m$ |
| 4 | Competitive | Adjacent gap or backward gap (Initial decision: non-cooperative) | $85\ km/h$ | $72\ km/h$ | $14.0\ m$ |
| 5 | | Adjacent gap or backward gap (Initial decision: cooperative) | $90\ km/h$ | $75\ km/h$ | $7.5\ m$ |

*6.4. Case study results*

Cooperative and competitive cases were tested using the developed simulation model. In order to validate the repeated game model's performance, the simulation results using the repeated model are compared with results using the calibrated stage game model played independently, i.e., one-shot game model at every decision-making point.

In cooperative scenarios, a dominant action strategy is found in rational decision-making due to the apparent situation. The simulation model using the repeated game model shows very close performance with that using the one-shot game as the game results are same in each game point. Since there is a mixed strategy Nash equilibrium in the competitive cases, both drivers decide an action strategy depending on the probability of actions. For case study results, this study provides the typical outcome of each scenario if there is no distinct difference in decision-making using the two game models. Otherwise, especially in the competitive scenario, the decision-making output simulation results of each game model are individually presented.

### 6.4.1. Case 1: cooperative merging scenario using an adjacent gap

As described in the sensitivity analysis, the developed game model has the ability to represent drivers' decisions in normal cooperative merging cases. According to the game results, as shown in Figure 17, drivers chose a 'change ($s_1$) and yield ($l_1$)' action set during the game period. The SV slightly accelerated by speed harmonization rules in preparation for merging while the LV decelerated in order to accept the SV's lane change. When a lead and lag gap was acceptable, the SV merged onto the freeway mainline. In simulation, the driver of SV controlled the vehicle's speed via the car-following rule as soon as executing the lane change and its following vehicles also showed oscillation in their speed profiles to ensure a safe gap.

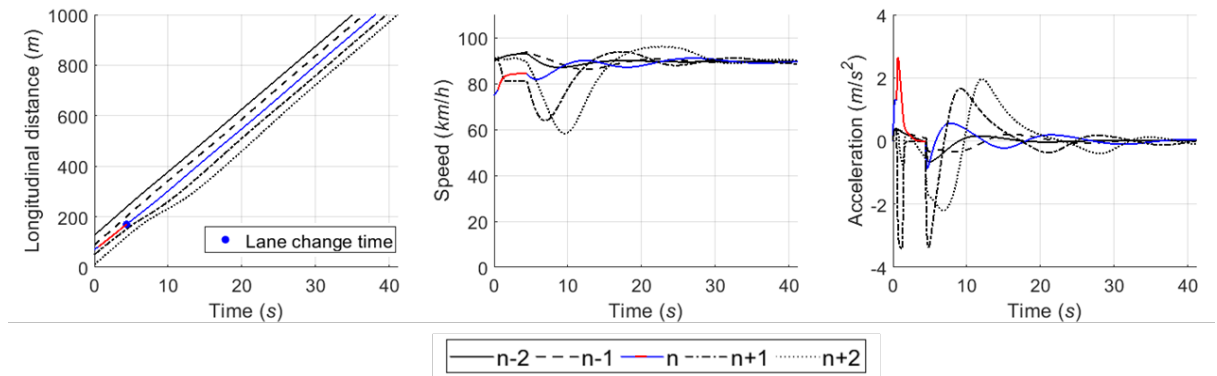

Note that a red solid line indicates simulation data of the SV (vehicle $n$) during game period, whereas a blue solid line shows the SV's data in simulation time except game period.

**Figure 16.** Graphical representation of simulation results in case 1.

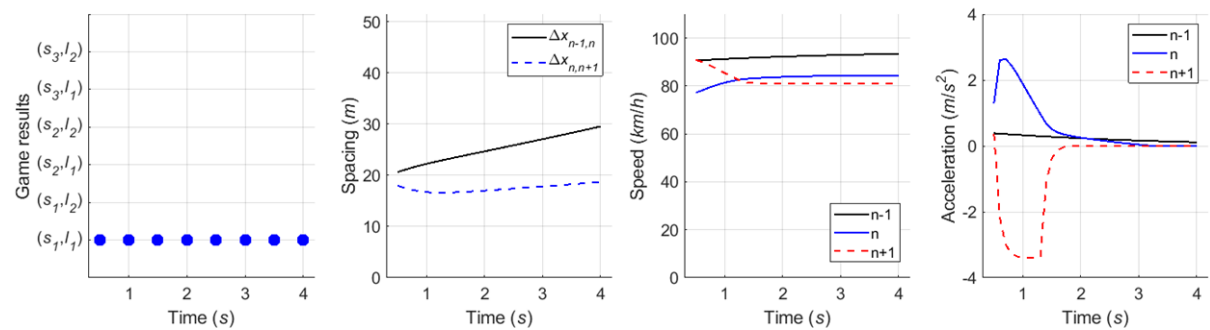

**Figure 17.** Decision-making game results in case 1.

### 6.4.2. Case 2: cooperative merging scenario using a backward gap

Simulation results for the second case, as shown in Figure 18, indicate that the driver of SV used the backward gap after the initial LV to overtake the SV. In Figure 19(a), the drivers decided on a 'wait ($s_2$) and block ($l_2$)' action strategy, respectively. The LV accelerated to block merging, and the SV also accelerated for speed synchronization even though the driver of SV decided to take a 'wait ($s_2$)' action. As soon as the initial LV overtook the SV, a new merging decision-making game was identified in which the vehicle $n + 2$ became the new LV. The results of the second game are shown in Figure 19(b). The SV continuously chose a 'change ($s_1$)' action until the gap acceptance rule was satisfied, then moved to the freeway mainline in consideration of gap size and relative speed. The LV, i.e., the vehicle $n + 2$, in the second game decelerated in a yielding action in response to the SV's intention to merge. In conclusion, the merging decision-making model was shown to depict a typical waiting scenario for both game models.

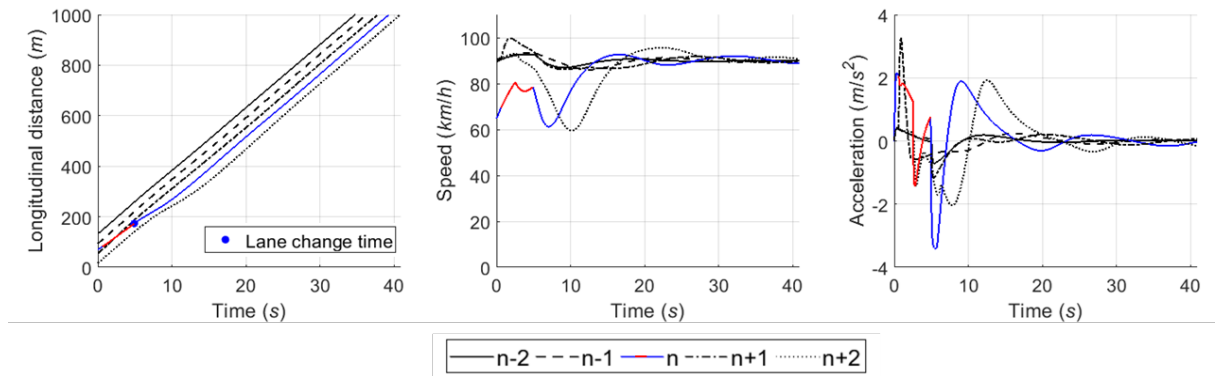

Note that a red solid line indicates simulation data of the SV (vehicle $n$) during game period, whereas a blue solid line shows the SV's data in simulation time except game period.

**Figure 18.** Graphical representation of simulation results in case 2.

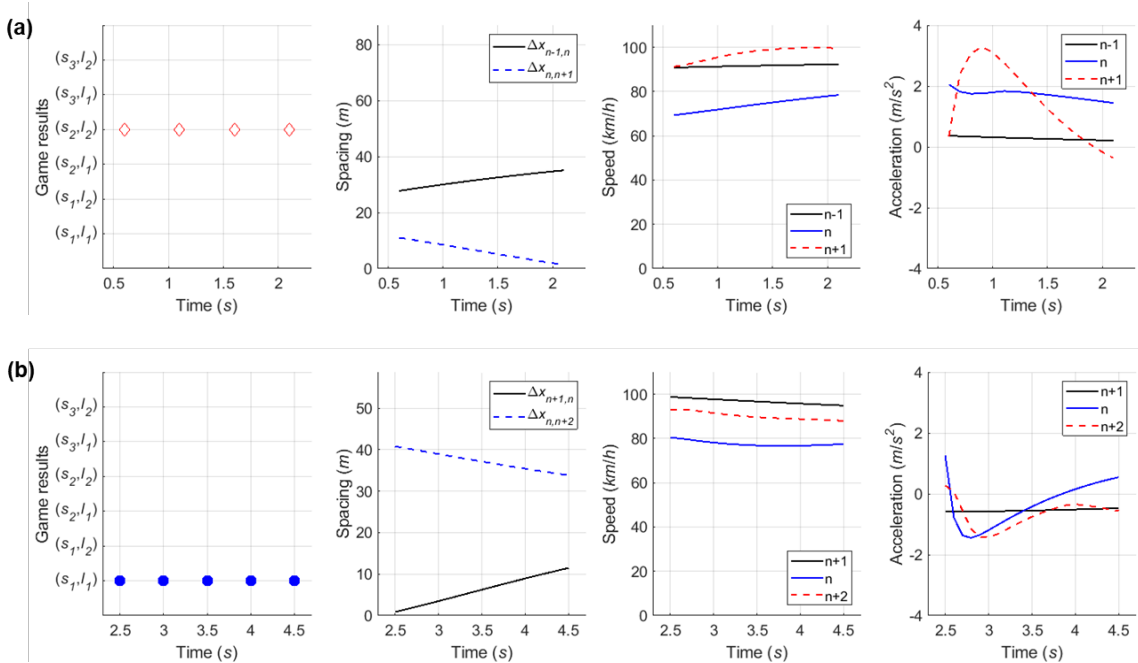

**Figure 19.** Decision-making game results in case 2.

### 6.4.3. Case 3: cooperative merging scenario using a forward gap

In overtaking scenario, time-space diagram in Figure 20 shows that the SV took the forward gap and then merge onto the freeway. When the SV entered the acceleration lane, the SV and LV chosen the 'overtake ($s_3$) and yield ($l_1$)' action set. Although the LV decided the yielding action, it was observed that the LV maintained its speed during the first game period due to observing the SV's passing. After overtaking the lead vehicle, the SV began to decrease the speed to harmonize with that of freeway vehicles. New LV, i.e., it had been the lead vehicle in the first game period, selected the yielding action in interaction with the SV. So it showed the deep deceleration during the second game period. The SV maintained on the acceleration lane, then it changed a lane as soon as the gap acceptance rule was satisfied. As described in simulation setting, overtaking scenario is usually observed in congested traffic condition. Thus this lane-changing by overtaking action caused huge oscillation in speed profile because generally spacing between vehicles is small under congested traffic condition. It is concluded that this simulation model based on the proposed game model well represents inducing a backward forming shockwave by merging traffic in congested condition.

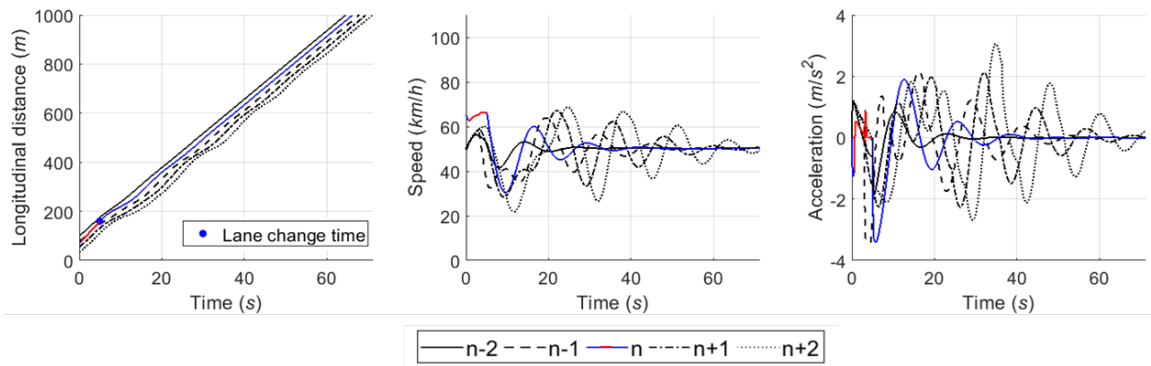

Note that a red solid line indicates simulation data of the SV (vehicle $n$) during game period, whereas a blue solid line shows the SV's data in simulation time except game period.

**Figure 20.** Graphical representation of simulation results in case 3.

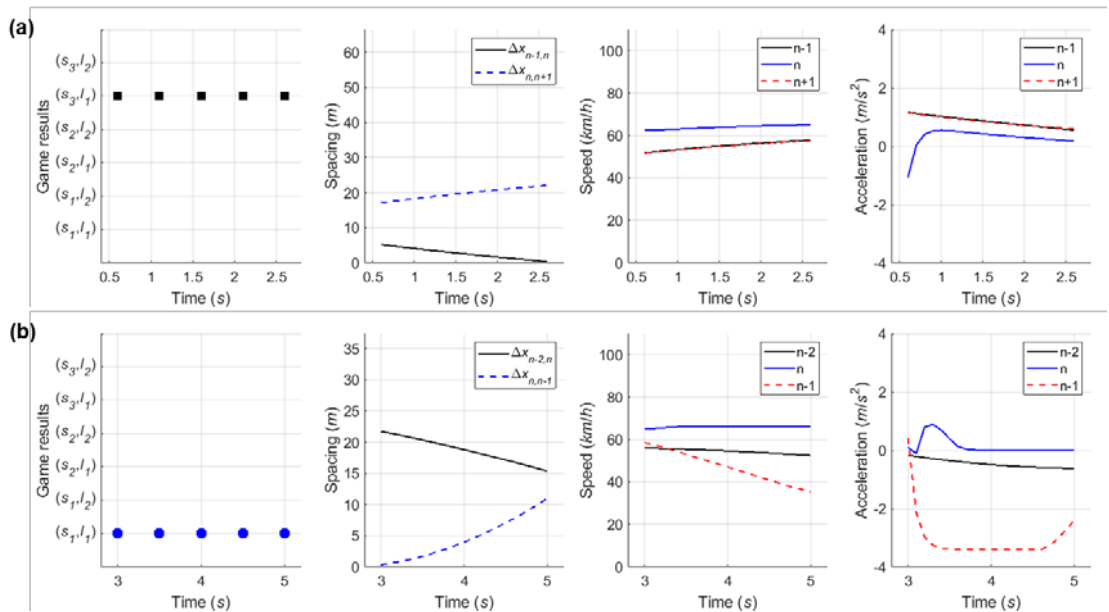

**Figure 21.** Decision-making game results in case 3.

### 6.4.4. Case 4: competitive merging scenario choosing an adjacent gap or a backward gap (1)

In the fourth competitive merging case, the initial game result of ($s_1, l_2$) is observed in Figure 23(a). As a non-cooperative action strategy set, it means that both drivers are in competition to achieve their own objective. At the third decision-making point, a decision they make becomes ($s_2, l_2$) as a cooperative action strategy set. Although the driver of SV initially wanted to change a lane using an adjacent gap as soon as entering an acceleration lane, they change the initial decision in order to avoid collision after recognizing the opposite driver's aggressive behavior. Thus the driver finally uses the backward gap for merging onto the freeway. By this case, this study conclude that the repeated game model enables to depict practical change of drivers' decision in competitive decision-making even using the cumulative function.

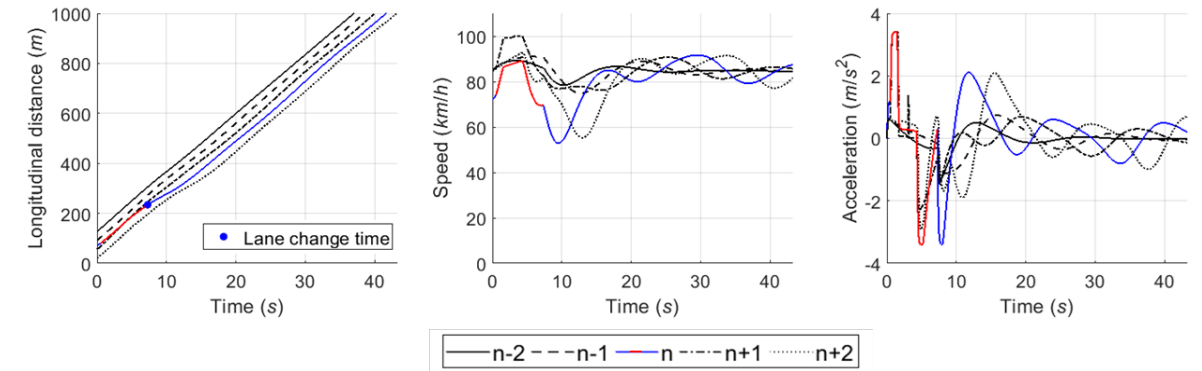

Note that a red solid line indicates simulation data of the SV (vehicle *n*) during game period, whereas a blue solid line shows the SV's data in simulation time except game period.

**Figure 22.** Graphical representation of simulation results in case 4.

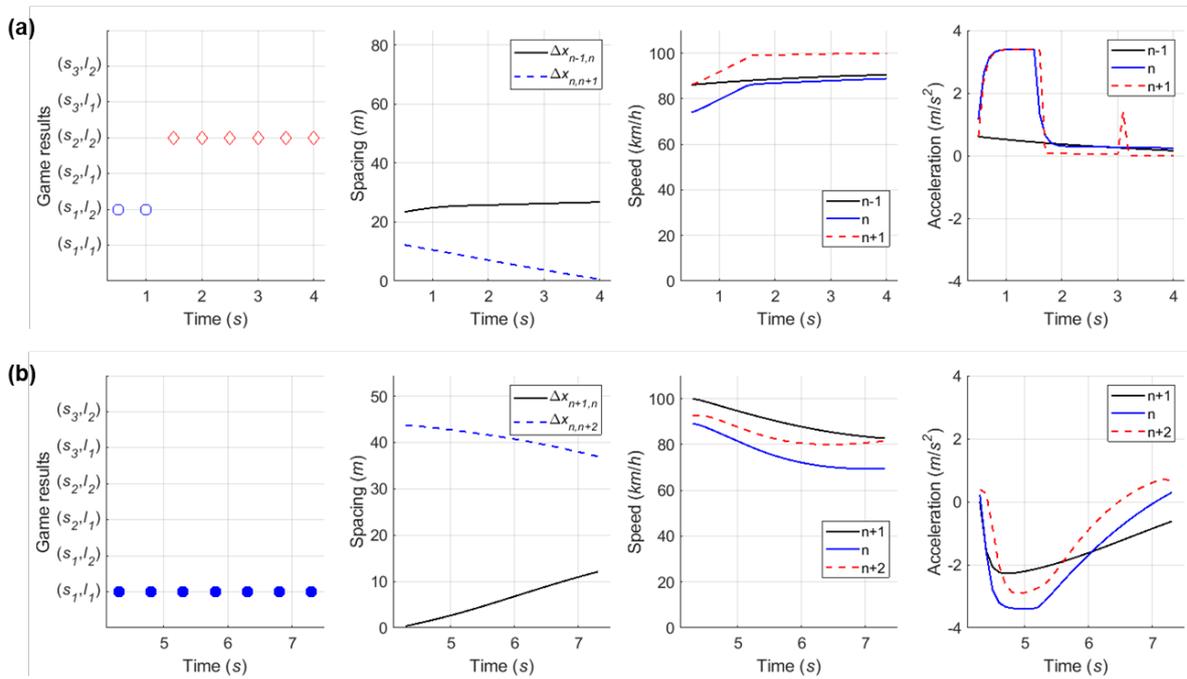

**Figure 23.** Decision-making game results in case 4.

### 6.4.5. Case 5: competitive merging scenario choosing an adjacent gap or a backward gap (2)

In Case 5, the simulation results show the SV used the backward gap for merging onto the freeway whichever game model is used, as illustrated in Figure 24 and Figure 25. This example shows competition to choose an adjacent gap or a backward gap, as in Case 4. However, there is a difference in that the initial decision is a cooperative action strategy in Case 5.

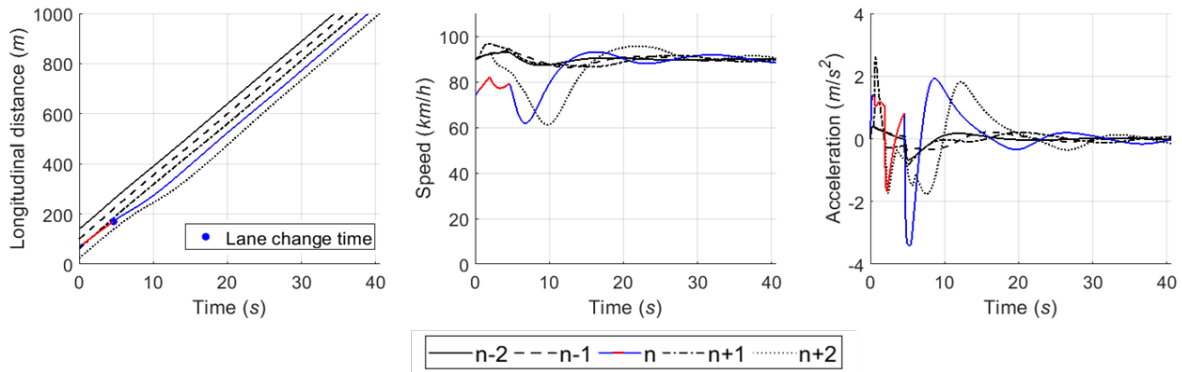

Note that a red solid line indicates simulation data of the SV (vehicle *n*) during game period, whereas a blue solid line shows the SV's data in simulation time except game period.

**Figure 24.** Graphical representation of simulation results in case 5 using the repeated game model.

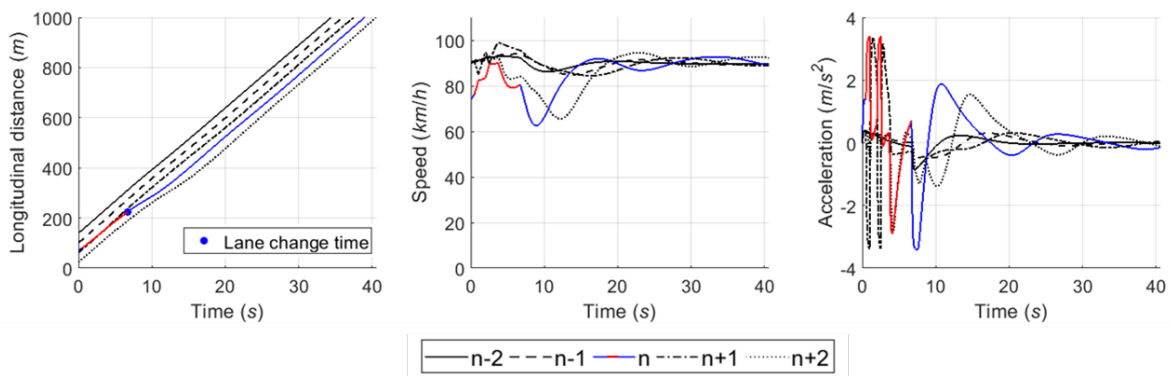

Note that a red solid line indicates simulation data of the SV (vehicle *n*) during game period, whereas a blue solid line shows the SV's data in simulation time except game period.

**Figure 25.** Graphical representation of simulation results in case 5 using the one-shot game model.

In Figure 26(a), when the repeated game model was used, the driver of SV chose a 'wait $(s_2)$' action during the first game period and then decided to change a lane in the second game period. While decision-making results were maintained using the repeated game model, oscillation in decision-making is revealed when the one-shot game is used, as shown in Figure 27(a). One reason why the one-shot game model causes unstable decision results is that the stage game decides a driver's action in a merging situation based on instantaneous vehicle location, speed, and acceleration data without consideration of previous game results (i.e., decisions made at previous game points). Considering the goal of each action, a change from a non-cooperative strategy set to a cooperative strategy is required in order to avoid a collision (if $(s_1, l_2)$ is chosen) or unnecessary deceleration (if $(s_2, l_1)$ is selected). However, changes between cooperative action strategy sets (i.e., $(s_1, l_1)$ and $(s_2, l_2)$) are not realistic except when there is a surrounding vehicle intervention. This case shows a distinct difference observed in simulation results depending on which type of the two game models is used. Oscillation in decision-making may reduce the performance of microscopic traffic simulation models even though it is only observed in specific competitive merging situations.

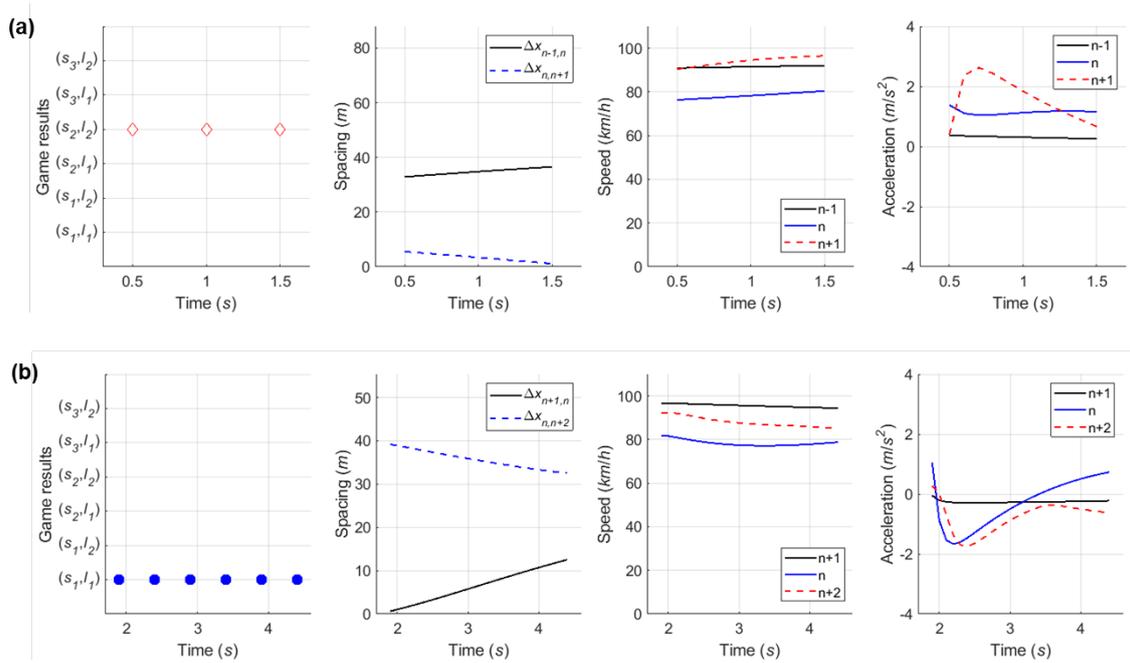

**Figure 26.** Decision-making game results in case 5 using the repeated game model.

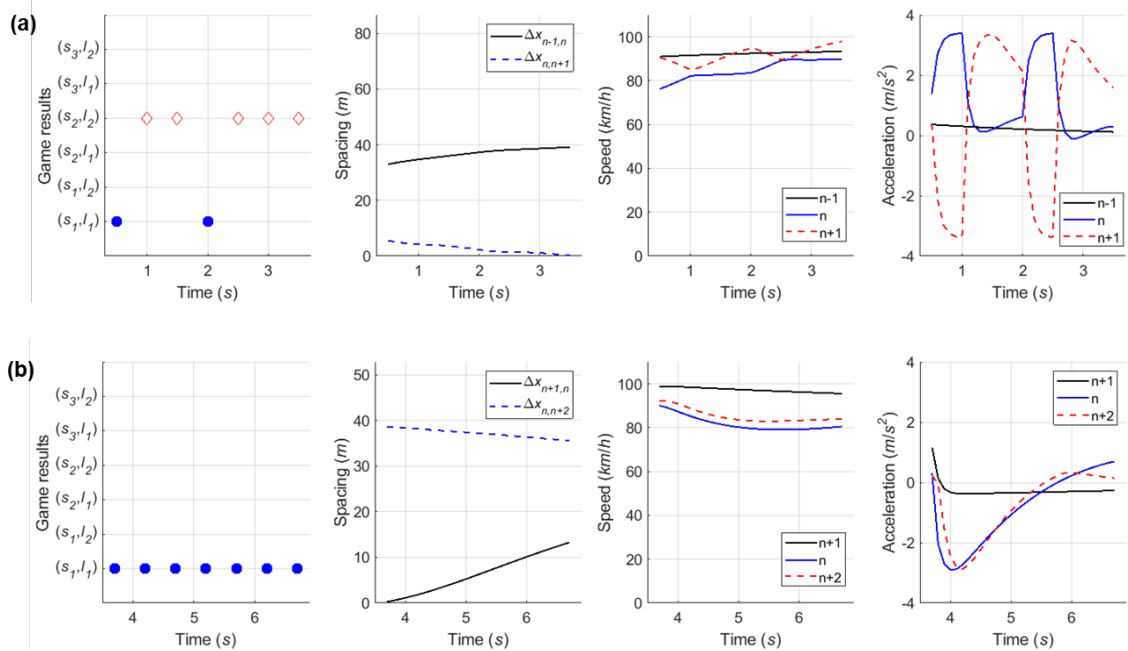

**Figure 27.** Decision-making game results in case 5 using the one-shot game model.

## 7. Conclusions

Drivers' behavior has a big impact on the safety and throughput of the transportation system. This is especially true for traffic conflicts between merging and through vehicles, in that merging vehicles induce shockwaves, which results in a reduction in the roadway capacity resulting in traffic congestion. Consequently, modeling driving behavior thoroughly and accurately is critical to analyzing traffic flow in microscopic traffic simulation and to taking advantage of the advanced vehicle driving technologies and strategies in AVs. The purpose of this study is to update the repeated game lane-changing model proposed in [2]. This game model has a feature that interprets interaction between drivers, as compared to most lane-changing models, which are focused on the lane-changing vehicle only. In this study, the payoff functions were newly formulated focusing on not only improvements in prediction performance but also use in microscopic traffic simulators. In the model evaluation, the developed model captured drivers' merging behaviors with a prediction accuracy of about 86%, showing improvement of about 12% compared to [2]. Also, this study presented the sensitivity analysis to indicate that the developed model can depict rational merging decision-making according to variations of the related factors: game location, relative speed, and gap size. In order to demonstrate why the repeated game is required in microscopic traffic simulation, moreover, a case study was conducted using the ABM developed to simulate merging situations. Using the repeated game model showed that it had superior performance to a one-shot game model, in which the stage game is independently played, in terms of representing practical merging behaviors in cooperative and competitive merging scenarios.

In order to elaborate on this study as a state-of-the-art lane-changing model, the decision-making model based on the game theoretical approach needs to be expanded as a decision-making model for both mandatory and discretionary lane changing. Since lane-changing-related decision making can be affected by several factors (e.g., road design, traffic stream condition, driving skill, driver's aggressiveness, etc.), the model should be calibrated based on field data collected in various conditions. Lastly, the game model can be applied to advanced vehicle systems, such as AVs, which coexist with human-operated vehicles on the roadway. The model based on the game theoretical approach is anticipated to become an appropriate model to decide lane-changing maneuvers and predict surrounding vehicle drivers' behaviors.

**Author Contributions:** conceptualization, K.K.; methodology, K.K. and H.A.R; validation, K.K.; simulation, K.K.; formal analysis, K.K. and H.A.R; writing—original draft preparation, K.K.; writing—review and editing, H.A.R.; visualization, K.K.; supervision, H.A.R.

**Funding:** This research was funded partially by the University Mobility and Equity Center (UMEC) and a gift from the Toyota InfoTechnology Center.

**Conflicts of Interest:** The authors do not have any conflict of interest with other entities or researchers.